\documentclass[twocolumn,trackchanges]{aastex631}
\usepackage{etoolbox}          

\AtBeginEnvironment{figure}{\nolinenumbers}
\AtEndEnvironment{figure}{\linenumbers}
\AtBeginEnvironment{figure*}{\nolinenumbers}
\AtEndEnvironment{figure*}{\linenumbers}
\AtBeginEnvironment{table}{\nolinenumbers}
\AtEndEnvironment{table}{\linenumbers}
\AtBeginEnvironment{table*}{\nolinenumbers}
\AtEndEnvironment{table*}{\linenumbers}
\AtBeginEnvironment{deluxetable}{\nolinenumbers}
\AtEndEnvironment{deluxetable}{\linenumbers}
\AtBeginEnvironment{deluxetable*}{\nolinenumbers}
\AtEndEnvironment{deluxetable*}{\linenumbers}

\newcommand*\patchAmsMathEnvironmentForLineno[1]{%
  \expandafter\pretocmd\csname #1\endcsname{\linenomath}{}{}%
  \expandafter\pretocmd\csname end#1\endcsname{\endlinenomath}{}{}%
}
\newcommand*\patchBothAmsMathEnvironmentsForLineno[1]{%
  \patchAmsMathEnvironmentForLineno{#1}%
  \patchAmsMathEnvironmentForLineno{#1*}%
}
\AtBeginDocument{%
  \patchBothAmsMathEnvironmentsForLineno{equation}%
  \patchBothAmsMathEnvironmentsForLineno{align}%
  \patchBothAmsMathEnvironmentsForLineno{flalign}%
  \patchBothAmsMathEnvironmentsForLineno{alignat}%
  \patchBothAmsMathEnvironmentsForLineno{gather}%
  \patchBothAmsMathEnvironmentsForLineno{multline}%
}
\makeatletter
\let\frontmatter@title@above=\relax
\makeatother

\usepackage{subfig}
\usepackage{url}
\usepackage{amsmath}
\usepackage{soul,xcolor}
\usepackage{ amssymb }
\usepackage{appendix}
\usepackage{float}
\usepackage{lipsum}

\shorttitle{Hyperonic Direct-Urca Neutrino Emission with Lattice-QCD-Informed Form Factors}

\graphicspath{Figures/}
    \usepackage{graphicx}
    \usepackage{capt-of}
    \usepackage{lipsum}

\begin{document}

\setstcolor{red}
\correspondingauthor{}
\email{andreas.konstandinu@gmail.com}

\author[0000-0002-1072-7313]{Andreas Konstantinou}
\affiliation{Department of Physics, University of Cyprus, P.O. Box 20537, 1678 Nicosia, Cyprus}
\affiliation{Computation-based Science and Technology Research Center,
The Cyprus Institute, 20 Kavafi Str., Nicosia 2121, Cyprus}

\title{Hyperonic Direct-Urca Neutrino Emission with Lattice-QCD-Informed Form Factors}

\accepted{by ApJ}
\begin{abstract}
We investigate the sensitivity of direct-Urca neutrino emission in hyperonic neutron-star matter to the structure and momentum dependence of baryonic transition form factors. Using the zero-temperature FSU2H relativistic mean-field equation of state, we derive the relativistic emissivity for general spin-$1/2$ baryon transitions with the complete six form-factor contribution. We compare conventional static flavor-SU(3) couplings with vacuum lattice-QCD-informed form factors, using direct lattice vacuum input for $n\rightarrow p$ and $\Lambda\rightarrow p$ and a hybrid construction for the remaining hyperonic channels. Evaluating the latter both at $q^2=0$ and at the kinetic momentum transfer sampled at the Fermi surfaces separates changes in the zero-momentum couplings from momentum-transfer effects. The sensitivity is strongly channel dependent. The nucleonic transition probes $q^2\simeq0$ and is only weakly affected, whereas hyperonic channels can sample substantially larger momentum transfers and receive correspondingly larger corrections. These modifications propagate to the integrated neutrino luminosity of stellar configurations. Our results estimate the possible scale of the sensitivity coming from momentum transfer dependence, absent from static-coupling treatments and motivate future first-principles studies of baryonic weak currents in dense matter.
\end{abstract}


\section{Introduction}
\label{sec:introduction}

The thermal evolution of an isolated neutron star is governed by its
heat capacity, thermal transport, photon emission from the surface,
and neutrino emission from the stellar interior
\citep{Yakovlev:2000jp,Yakovlev:2004iq,Potekhin:2015qsa}.
During the neutrino-cooling era, neutrino emission from the stellar
interior provides the dominant energy-loss mechanism.
Direct-Urca reactions are particularly efficient, with an emissivity
scaling as $T^6$, rather than the $T^8$ behavior characteristic of
modified-Urca processes
\citep{Lattimer:1991ib,Yakovlev:2000jp,Yakovlev:2004iq}.
The onset of a direct-Urca channel can therefore qualitatively alter the cooling history of a neutron star and provide a sensitive probe of the composition of matter at supranuclear densities \citep{Lattimer:1991ib,Yakovlev:2004iq}.

At baryon densities of a few times nuclear saturation density,
hyperons may become energetically favored, depending sensitively on
the hyperon interactions and the adopted equation of state (EOS)
\citep{1960SvA.....4..187A,Glendenning:1982nc,
Chatterjee:2015pua,Tolos:2017lgv}.
Their appearance modifies the composition of beta-equilibrated matter
and opens additional charged-current direct-Urca reactions,
\begin{equation}
    B_1 \rightarrow B_2+\ell^-+\bar{\nu}_{\ell},
    \qquad
    B_2+\ell^- \rightarrow B_1+\nu_{\ell}.
\end{equation}
Such reactions have long been recognized as possible mechanisms for
rapid neutron-star cooling \citep{Prakash:1992zng}, and calculations
with the hyperonic FSU2H EOS have shown that they can substantially
affect the thermal evolution of sufficiently massive stars
\citep{Tolos:2019wea}.
Their contribution depends not only on the EOS and the resulting
particle composition, but also on the weak baryonic transition vertex.

For semileptonic transitions between spin-$1/2$ baryons, the vacuum
charged-current matrix element is described by six form factors:
the vector, weak-magnetism, and induced-scalar form factors
$f_1$, $f_2$, and $f_3$, and the axial, weak-electricity, and
induced-pseudoscalar form factors $g_1$, $g_2$, and $g_3$
\citep{Gaillard:1984ny,Cabibbo:2003cu}.
Hyperonic direct-Urca calculations have commonly approximated these
vertices using momentum-independent vector and axial couplings related
through flavor SU(3), with the remaining structures neglected or
treated phenomenologically
\citep{Prakash:1992zng,Yakovlev:2000jp}.
These approximations neglect, to different degrees, the
momentum-transfer and density dependence of the form factors, flavor-SU(3)
breaking, and contributions from subleading structures of the weak
current.

Lattice QCD now provides first-principles information on baryonic
form factors and their momentum-transfer dependence in vacuum.
For the nucleon, electromagnetic and axial form factors have been
calculated at or near the physical pion mass, including continuum
extrapolations
\citep{Alexandrou:2023qbg,Alexandrou:2025vto}.
Lattice calculations have also quantified flavor-SU(3) breaking in
hyperon semileptonic transitions, including studies of
$\Sigma^-\rightarrow n$ and $\Xi^0\rightarrow\Sigma^+$
\citep{Guadagnoli:2006gj,Sasaki:2008ha,Sasaki:2012ne,Sasaki:2017jue}.
More recently, all six vector and axial-vector form factors for the
$\Lambda\rightarrow p$ transition, including the second-class form
factors, have been determined directly at the physical pion mass at one lattice spacing
\citep{Bacchio:2025auj}.

Applying vacuum form factors to baryonic quasiparticles in dense
matter nevertheless requires care.
The medium introduces a preferred four-velocity $u^\mu$, so that the
complete in-medium weak vertex may contain additional structures and
depend separately on $q^2$, $q\cdot u$, and the thermodynamic state of
the matter \citep{Weldon:1982aq,Leinson:2002bw}.
A first-principles lattice-QCD determination of this vertex at the
large real baryon chemical potentials relevant to neutron-star cores
is presently prevented by the finite-density sign problem
\citep{deForcrand:2009zkb}.
The use of vacuum lattice-QCD form factors should therefore not be
interpreted as a determination of the complete in-medium weak current.
Instead, it provides a controlled way to quantify the sensitivity of
direct-Urca rates to the known vacuum momentum-transfer dependence.

In this work, we study this sensitivity for relativistic direct-Urca
reactions in beta-equilibrated matter described by the
zero-temperature FSU2H relativistic mean-field EOS
\citep{Tolos:2017lgv,Tolos:2016hhl}.
We evaluate the emissivity using the complete six-form-factors, retaining all quadratic and interference contributions.
A conventional static flavor-SU(3) description at $q^2=0$ is used as
the reference.
We then employ direct lattice-QCD input for the
$n\rightarrow p$ and $\Lambda\rightarrow p$ transitions and a hybrid
SU(3)--lattice construction for the remaining hyperonic channels.
By evaluating the lattice-informed form factors both at $q^2=0$ and
at the kinetic momentum transfer determined by the Fermi-surface
kinematics, we separate changes in the zero-momentum form-factor input
from momentum-transfer effects.
We examine their impact on the density-dependent direct-Urca
emissivities and on the redshifted neutrino luminosity of FSU2H
stellar configurations.

The paper is organized as follows.
Section~\ref{sec:eos} introduces the FSU2H equation of state.
Section~\ref{sec:current} derives the relativistic direct-Urca emissivity with the complete six-form-factor contribution.
Section~\ref{sec:form_factor_inputs} describes the static SU(3), direct lattice-QCD, and hybrid form-factor inputs.
Section~\ref{sec:results} presents the density-dependent emissivities and stellar neutrino luminosities.
We summarize our conclusions in Section~\ref{sec:conclusions}.

\section{Equation of State}
\label{sec:eos}

The neutrino emissivities are evaluated using the zero-temperature, zero-magnetic-field FSU2H relativistic mean-field (RMF) equation of state of \citet{Tolos:2017lgv,Tolos:2016hhl}. In this framework, baryons interact through the exchange of scalar and vector mesons, while electrons and muons are treated as free relativistic Fermi gases. The model contains the complete baryon octet,
\begin{equation}
    B\in
    \left\{
    n,p,\Lambda,
    \Sigma^{-},\Sigma^{0},\Sigma^{+},
    \Xi^{-},\Xi^{0}
    \right\},
\end{equation}
together with electrons and muons in beta-equilibrated, charge-neutral matter. Throughout this work we set the magnetic field to zero, so that the electromagnetic field contributions and Landau quantization present in the general formulation are absent. 

The RMF Lagrangian density is
\citep{osti_5443763,Serot:1997xg,2000csnp.conf.....G,Chen:2014sca}
\begin{equation}
    \mathcal{L}
    =
    \sum_B \mathcal{L}_B
    +
    \mathcal{L}_m
    +
    \sum_{l=e,\mu}\mathcal{L}_l .
    \label{eq:total_lagrangian}
\end{equation}
The baryon and lepton contributions are
\begin{align}
\mathcal{L}_B
={}&
\bar{\psi}_B
\bigg[
\gamma_\mu
\left(
i\partial^\mu
-g_{\omega B}\omega^\mu
-g_{\rho B}\bm{I_B}\cdot\bm{\rho}^{\,\mu}
-g_{\phi B}\phi^\mu
\right)\nonumber\\
&-\left(M_B-g_{\sigma B}\sigma\right)
\bigg]
\psi_B ,
\label{eq:lag_baryon}
\\
\mathcal{L}_l
={}&
\bar{\psi}_l
\left(
i\gamma_\mu\partial^\mu-m_l
\right)
\psi_l .
\label{eq:lag_lepton}
\end{align}
Here, \(\sigma\) denotes the scalar-isoscalar meson,
\(\omega^\mu\) the vector-isoscalar meson,
\(\bm{\rho}^{\,\mu}\) the vector-isovector meson, and
\(\phi^\mu\) the hidden-strangeness vector meson.
The corresponding baryon--meson coupling constants are denoted by
\(g_{\sigma B}\), \(g_{\omega B}\), \(g_{\rho B}\), and
\(g_{\phi B}\), while \(\bm{I}_B\) is the isospin operator of baryon
\(B\).

The mesonic contribution is
\begin{align}
\mathcal{L}_m
={}&
\frac{1}{2}
\partial_\mu\sigma\,\partial^\mu\sigma
-\frac{1}{2}m_\sigma^2\sigma^2
-\frac{\kappa}{3!}
\left(g_{\sigma N}\sigma\right)^3
-\frac{\lambda}{4!}
\left(g_{\sigma N}\sigma\right)^4
\nonumber\\
&-
\frac{1}{4}
\Omega_{\mu\nu}\Omega^{\mu\nu}
+
\frac{1}{2}m_\omega^2\omega_\mu\omega^\mu
+
\frac{\zeta}{4!}
g_{\omega N}^4
\left(\omega_\mu\omega^\mu\right)^2
\nonumber\\
&-
\frac{1}{4}
\bm{R}_{\mu\nu}\cdot\bm{R}^{\mu\nu}
+
\frac{1}{2}m_\rho^2
\bm{\rho}_\mu\cdot\bm{\rho}^{\,\mu}
\nonumber\\
&+
\Lambda_\omega
g_{\rho N}^2g_{\omega N}^2
\left(
\bm{\rho}_\mu\cdot\bm{\rho}^{\,\mu}
\right)
\left(
\omega_\nu\omega^\nu
\right)
\nonumber\\
&-
\frac{1}{4}
P_{\mu\nu}P^{\mu\nu}
+
\frac{1}{2}m_\phi^2\phi_\mu\phi^\mu ,
\label{eq:lag_meson}
\end{align}
where the meson field-strength tensors are
\begin{align}
    \Omega_{\mu\nu}
    &=
    \partial_\mu\omega_\nu-\partial_\nu\omega_\mu ,
    \\
    \bm{R}_{\mu\nu}
    &=
    \partial_\mu\bm{\rho}_\nu
    -
    \partial_\nu\bm{\rho}_\mu ,
    \\
    P_{\mu\nu}
    &=
    \partial_\mu\phi_\nu-\partial_\nu\phi_\mu .
\end{align}
The nonlinear scalar, isoscalar-vector, and mixed isoscalar-isovector interactions generate a nonlinear density dependence of the self-consistent meson mean fields and, consequently, of the effective baryon masses, vector potentials, particle composition, and bulk EOS. The coupling constants appearing in the Lagrangian remain fixed and are not promoted to explicitly density-dependent couplings.

In the mean-field approximation, the meson fields are replaced by
their expectation values. For uniform and isotropic matter, only the
time components of the vector fields remain nonzero
\begin{equation}
    \sigma\rightarrow\bar{\sigma},
    \
    \omega^\mu\rightarrow
    \delta^{\mu 0}\bar{\omega},
    \ 
    \bm{\rho}^{\,\mu}\rightarrow
    \delta^{\mu 0}\bar{\rho}\,\hat{\bm e}_3,
    \ 
    \phi^\mu\rightarrow
    \delta^{\mu 0}\bar{\phi}.
    \label{eq:mean_fields}
\end{equation}
The scalar field generates the effective baryon mass
\begin{equation}
    M_B^*
    =
    M_B-g_{\sigma B}\bar{\sigma},
    \label{eq:effective_mass}
\end{equation}
whereas the vector fields generate the single-particle potential
\begin{equation}
    U_B
    =
    g_{\omega B}\bar{\omega}
    +
    g_{\rho B}I_{3B}\bar{\rho}
    +
    g_{\phi B}\bar{\phi}.
    \label{eq:vector_potential}
\end{equation}
Here, \(I_{3B}\) is the third component of the baryon isospin, with
the convention \(I_{3p}=+1/2\) and \(I_{3n}=-1/2\).

At zero temperature, the kinetic Fermi energy and chemical potential
of baryon \(B\) are
\begin{align}
    \epsilon_{FB}
    &=
    \sqrt{k_{FB}^2+M_B^{*2}},
    \label{eq:kinetic_fermi_energy}
    \\
    \mu_B
    &=
    \epsilon_{FB}+U_B ,
    \label{eq:chem_baryon}
\end{align}
where \(k_{FB}\) is the baryon Fermi momentum. For leptons,
\begin{equation}
    \mu_l
    =
    \sqrt{k_{Fl}^2+m_l^2},
    \qquad
    l=e,\mu .
    \label{eq:chem_lepton}
\end{equation}
The number densities are related to the corresponding Fermi momenta
by
\begin{equation}
    n_i
    =
    \frac{k_{Fi}^3}{3\pi^2},
    \label{eq:number_density}
\end{equation}
for each spin-\(1/2\) fermion species \(i\).

The mean fields satisfy
\begin{align}
m_\sigma^2\bar{\sigma}&
+
\frac{\kappa}{2}
g_{\sigma N}^3\bar{\sigma}^{\,2}
+
\frac{\lambda}{3!}
g_{\sigma N}^4\bar{\sigma}^{\,3}
=
\sum_B g_{\sigma B}n_B^s ,
\label{eq:sigma_field}
\\
m_\omega^2\bar{\omega}&
+
\frac{\zeta}{3!}
g_{\omega N}^4\bar{\omega}^{\,3}
+
2\Lambda_\omega
g_{\rho N}^2g_{\omega N}^2
\bar{\rho}^{\,2}\bar{\omega}
=
\sum_B g_{\omega B}n_B ,
\label{eq:omega_field}
\\
m_\rho^2\bar{\rho}&
+
2\Lambda_\omega
g_{\rho N}^2g_{\omega N}^2
\bar{\omega}^{\,2}\bar{\rho}
=
\sum_B g_{\rho B}I_{3B}n_B ,
\label{eq:rho_field}
\\
m_\phi^2\bar{\phi}
&=
\sum_B g_{\phi B}n_B ,
\label{eq:phi_field}
\end{align}
where the baryon scalar density is
\begin{equation}
n_B^s
=
\frac{M_B^*}{2\pi^2}
\left[
k_{FB}\epsilon_{FB}
-
M_B^{*2}
\ln\left(
\frac{k_{FB}+\epsilon_{FB}}{M_B^*}
\right)
\right].
\label{eq:scalar_density}
\end{equation}

The matter energy density is obtained from the mean-field
energy--momentum tensor,
\begin{align}
\varepsilon
={}&
\sum_B \varepsilon_B+\sum_{l=e,\mu}\varepsilon_l\nonumber\\
&
+\frac{1}{2}m_\sigma^2\bar{\sigma}^2
+\frac{1}{2}m_\omega^2\bar{\omega}^2
+\frac{1}{2}m_\rho^2\bar{\rho}^2
+\frac{1}{2}m_\phi^2\bar{\phi}^2
\nonumber\\
&+
\frac{\kappa}{3!}(g_{\sigma N}\bar{\sigma})^3
+
\frac{\lambda}{4!}(g_{\sigma N}\bar{\sigma})^4
+
\frac{\zeta}{8}(g_{\omega N}\bar{\omega})^4
\nonumber\\
&+
3\Lambda_\omega
g_{\rho N}^2g_{\omega N}^2
\bar{\rho}^{\,2}\bar{\omega}^{\,2}.
\label{eq:energy_density}
\end{align}

At zero temperature the pressure follows from the thermodynamic
relation
\begin{equation}
P=\sum_i\mu_i n_i-\varepsilon .
\label{eq:pressure}
\end{equation}

Chemical equilibrium under weak interactions requires
\begin{equation}
    \mu_i
    =
    b_i\mu_n-q_i\mu_e ,
    \label{eq:beta_equilibrium}
\end{equation}
where \(b_i\) and \(q_i\) are the baryon number and electric charge of
species \(i\), respectively. In particular,
\begin{align}
    \mu_p
    &=\mu_{\Sigma^+}=
    \mu_n-\mu_e ,
    \nonumber\\
    \mu_\Lambda&
    =
    \mu_{\Sigma^0}
    =
    \mu_{\Xi^0}
    =
    \mu_n ,
    \nonumber\\
    \mu_{\Sigma^-}
    &=
    \mu_{\Xi^-}
    =
    \mu_n+\mu_e .
    \label{eq:beta_examples}
\end{align}
After muons appear,
\begin{equation}
    \mu_\mu=\mu_e .
    \label{eq:muon_equilibrium}
\end{equation}
Electric charge neutrality requires
\begin{equation}
    \sum_i q_i n_i=0 ,
    \label{eq:charge_neutrality}
\end{equation}
or explicitly,
\begin{equation}
    n_p+n_{\Sigma^+}
    =
    n_e+n_\mu+n_{\Sigma^-}+n_{\Xi^-}.
    \label{eq:charge_neutrality_explicit}
\end{equation}
The total baryon density is
\begin{equation}
    n_B
    =
    \sum_{i\in \text{baryons}} n_i .
    \label{eq:total_baryon_density}
\end{equation}

Equations
\eqref{eq:sigma_field}--\eqref{eq:total_baryon_density}
are solved self-consistently at fixed total baryon density. This
determines the particle fractions, effective masses, vector
potentials, chemical potentials, and Fermi momenta entering the
direct-Urca emissivity calculation. Throughout this work, we employ
the original zero-magnetic-field FSU2H parameter set of \citet{Tolos:2017lgv,Tolos:2016hhl}. For the FSU2H parameterization, the nuclear saturation density is
$n_0=0.1505~{\rm fm}^{-3}$.

For the stellar structure calculation, the FSU2H core EOS is joined to the crust EOS of \citet{Sharma:2015bna}, following the crust prescription adopted in the original FSU2H calculation \citep{Tolos:2017lgv}. The resulting complete EOS is used to solve the Tolman-Oppenheimer-Volkoff equations and obtain the radial density profiles employed in the neutrino-luminosity calculation. As a numerical validation, our implementation reproduces the
zero-magnetic-field FSU2H particle composition and stellar properties reported in ~\citet{Tolos:2017lgv,Tolos:2016hhl}. We obtain $M_{\rm max}\simeq2.03\,M_\odot$, a corresponding radius
of approximately $12.1\,{\rm km}$, a central density
$n_c\simeq5.8\,n_0$, and the onset of the first hyperon at
$n_B\simeq2.2\,n_0$, consistent with the published FSU2H results.

Although the neutrino emissivity is evaluated at finite temperature,
thermal corrections to the EOS and particle composition are neglected.
This is consistent with the strongly degenerate regime considered here, $T\ll\mu_i$, in which temperature enters primarily through the phase-space factors of the direct-Urca emissivity.

\section{Neutrino Emissivity}
\label{sec:current}

We consider the direct-Urca transition
\begin{equation}
    B_1(\mathcal{P}_1)
    \rightarrow
    B_2(\mathcal{P}_2)
    +
    \ell(k_\ell)
    +
        \bar{\nu}_\ell(k_{\nu_\ell}),
    \label{eq:reaction}
\end{equation}
where $B_1$ and $B_2$ denote the initial and final baryonic
quasiparticles, respectively, and $\ell=e,\mu$. The inverse
lepton-capture process occurs simultaneously in beta-equilibrated
matter. Neutrinos are assumed to escape freely, so their chemical
potential and final-state blocking factor are neglected. We use the
metric convention
$g_{\mu\nu}={\rm diag}(1,-1,-1,-1)$ and the gamma-matrix conventions
of \citet{Bjorken:1965zz}.

\subsection{Canonical and kinetic momentum transfer} \label{sec:q_vs_K}

An important distinction arises between canonical and kinetic
four-momenta in relativistic mean-field matter. For baryon species
$i$, the scalar mean field is incorporated through the effective mass
$M_i^*$, whereas the vector mean fields generate the single-particle
potential $U_i$. In uniform matter, the corresponding Dirac equation
can be written as
\begin{equation}
    \left[
        \gamma_\mu
        (
            i\partial^\mu-U_i u^\mu
        )
        -
        M_i^*
    \right]
    \psi_i
    =
    0,
    \qquad
    u^\mu=(1,\boldsymbol{0}),
    \label{eq:rmf_dirac_equation}
\end{equation}
where $u^\mu$ is the four-velocity of the medium.

For a stationary quasiparticle state with canonical four-momentum
$\mathcal{P}_i^\mu$, it is convenient to introduce the kinetic
four-momentum
\begin{equation}
    P_i^\mu
    \equiv
    \mathcal{P}_i^\mu-U_i u^\mu
    =
    \left(
        \epsilon_i,
        \boldsymbol{p}_i
    \right),
    \qquad
    \epsilon_i
    =
    \sqrt{
        \boldsymbol{p}_i^{\,2}
        +
        M_i^{*2}
    }.
    \label{eq:kinetic_momentum}
\end{equation}
The quasiparticle spinor then satisfies
\begin{equation}
    \left(
        \gamma_\mu P^\mu_i-M_i^*
    \right)
    u_i(P_i)
    =
    0.
    \label{eq:quasiparticle_dirac_equation}
\end{equation}
Thus, both scalar and vector mean-field interactions are retained. The scalar interaction modifies the baryon mass, while the vector
interaction relates the kinetic and canonical four-momenta through
\begin{equation}
    \mathcal{P}_i^\mu
    =
    P_i^\mu
    +
    U_i u^\mu.
    \label{eq:canonical_momentum}
\end{equation}

Total energy--momentum conservation applies to the canonical
four-momenta. The four-momentum transferred to the leptonic system is
therefore
\begin{equation}
    K^\mu
    \equiv
    \mathcal{P}_1^\mu-\mathcal{P}_2^\mu
    =
    k_\ell^\mu+k_{\nu_\ell}^\mu.
    \label{eq:canonical_transfer}
\end{equation}
The quantity $K^\mu$ consequently appears in the overall
energy--momentum conserving delta function of the reaction.

The kinetic four-momentum change between the initial and final
baryonic quasiparticles is instead
\begin{equation}
    q^\mu
    \equiv
    P_1^\mu-P_2^\mu.
    \label{eq:kinetic_transfer_definition}
\end{equation}
Equations~\eqref{eq:canonical_momentum}--\eqref{eq:kinetic_transfer_definition}
give
\begin{equation}
    q^\mu
    =
    K^\mu
    -
    \Delta U\,u^\mu,
    \qquad
    \Delta U
    \equiv
    U_1-U_2.
    \label{eq:q_K_relation}
\end{equation}

Following the relativistic mean-field quasiparticle treatment of
\citet{Leinson:2002bv}, we use the kinetic transfer $q^\mu$ in the
vacuum-like one-body baryonic vertex, while the canonical transfer
$K^\mu$ enforces total energy--momentum conservation and is carried
by the emitted lepton pair.

Notice that $K^\mu$ and $q^\mu$ coincide in vacuum, or whenever the two
baryons experience the same vector potential, but need not coincide
for transitions between different quasiparticle species in dense
matter.

The distinction is also related to the appearance of the preferred
medium four-velocity $u^\mu$. For example,
\begin{align}
    i\sigma^{\mu\nu}q_\nu
    &=
    i\sigma^{\mu\nu}K_\nu
    -
    \Delta U\,
    i\sigma^{\mu\nu}u_\nu,
    \label{eq:qK_tensor_relation}
    \\
    q^\mu
    &=
    K^\mu-\Delta U\,u^\mu.
    \label{eq:qK_vector_relation}
\end{align}
Consequently, rewriting a $q^\mu$-based quasiparticle current in
terms of $K^\mu$ generates explicit structures proportional to the
medium four-velocity. In a complete finite-density weak vertex these
structures may carry independent form factors
\citep{Weldon:1982aq,Leinson:2002bw}. The six-form-factor vertex
adopted below should therefore be regarded as the vacuum-like
one-body current associated with the RMF quasiparticle prescription,
rather than as the most general weak vertex allowed in the medium.

The distinction between the two momentum transfers becomes especially
transparent at the Fermi surfaces. In beta equilibrium,
\begin{equation}
    \mu_1-\mu_2=\mu_\ell,
    \label{eq:beta_equilibrium_transition}
\end{equation}
while the baryon chemical potentials satisfy
\begin{equation}
    \mu_i
    =
    \epsilon_{Fi}+U_i.
    \label{eq:mu_epsilon_U}
\end{equation}
It follows that
\begin{equation}
    \epsilon_{F1}-\epsilon_{F2}
    =
    \mu_\ell-\Delta U.
    \label{eq:kinetic_energy_transfer}
\end{equation}

In the strongly degenerate limit, the emitted antineutrino carries
energy and momentum of order $T$, much smaller than the Fermi
momenta of the baryons and charged leptons. Neglecting
$\boldsymbol{k}_{\nu_\ell}$ only when fixing the leading-order
Fermi-surface geometry gives
\begin{equation}
    \boldsymbol{q}
    =
    \boldsymbol{p}_1-\boldsymbol{p}_2
    \simeq
    \boldsymbol{k}_{F\ell}.
\end{equation}
The kinetic invariant entering the adopted baryonic vertex is
therefore
\begin{align}
    q_F^2
    &=
    \left(
        \epsilon_{F1}-\epsilon_{F2}
    \right)^2
    -
    k_{F\ell}^2
    \nonumber\\
    &=
    \left(
        \mu_\ell-\Delta U
    \right)^2
    -
    k_{F\ell}^2
    \nonumber\\
    &=
    m_\ell^2
    -
    2\mu_\ell\Delta U
    +
    \Delta U^2.
    \label{eq:q2_fermi_surface}
\end{align}
At the same order, the canonical momentum carried by the lepton pair
satisfies
\begin{align}
    K_F^2
    &\simeq
    \mu_\ell^2-k_{F\ell}^2
    \nonumber\\
    &=
    m_\ell^2.
    \label{eq:K2_fermi_surface}
\end{align}

Equations~\eqref{eq:q2_fermi_surface} and
\eqref{eq:K2_fermi_surface} show explicitly that the small invariant
mass of the emitted lepton pair does not imply a small kinetic
momentum transfer between the baryonic quasiparticles. The latter
also contains the difference in vector mean-field energies through
$\Delta U$. For transitions between different baryon species,
$\Delta U$ can be substantial at supranuclear densities, and
$q_F^2$ may therefore probe a hadronic momentum-transfer region very
different from that of the corresponding isolated semileptonic
decay.

This point is central to the interpretation of the form-factor
dependence considered below. Vacuum lattice-QCD calculations
determine baryonic matrix elements in the absence of the medium,
where the distinction between canonical and kinetic momenta is
absent. Evaluating these vacuum form-factor parametrizations at the
dense-matter kinetic transfer $q_F^2$ is therefore an embedding
prescription appropriate to the RMF quasiparticle current adopted
here. When $q_F^2$ lies outside the region directly constrained by
the corresponding vacuum decay or lattice-QCD calculation, this
requires an extrapolation of the vacuum momentum dependence.
Furthermore, genuine density-dependent modifications of the baryonic
weak vertex are not determined by the vacuum form factors. The
numerical calculations below should consequently be interpreted as
a sensitivity estimate of the possible scale of weak-vertex effects,
with an additional finite-density systematic uncertainty that cannot
presently be quantified.

\subsection{Weak vertex and direct-Urca rate}
\label{sec:weak_vertex}

The invariant amplitude is
\begin{align}
    \mathcal{M}_{12\ell}
    ={}&
    \frac{G_F C_{12}}{\sqrt{2}}
    \left[
        \bar{u}_\ell(k_\ell)
        \gamma_\mu (1-\gamma_5)
        v_\nu(k_{\nu_\ell})
    \right]
    \nonumber\\
    &\times
    \left[
        \bar{u}_2(P_2)
        \Gamma^\mu(q)
        u_1(P_1)
    \right],
    \label{eq:amplitude}
\end{align}
where $G_F$ is the Fermi constant and
\begin{equation}
    C_{12}
    =
    \begin{cases}
        V_{ud}, & \Delta S=0, \\[1mm]
        V_{us}, & |\Delta S|=1.
    \end{cases}
    \label{eq:cabibbo_factor}
\end{equation}
with \(\Delta S\) denoting the change in strangeness. 

In this work we use $G_F = 1.16638 \cdot (1 + 0.0105) \cdot 10^{-5}$\,GeV$^{-2}$ \citep{ParticleDataGroup:2024cfk}, where we account for radiative corrections by using the factor $(1 + 0.0105)$ ~\citep{Garcia:1985xz}. In the conventional Cabibbo notation, \(V_{ud}\simeq\cos\theta_C\) and \(V_{us}\simeq\sin\theta_C\). Numerically, we use \(V_{ud}=0.97420\) and \(V_{us}=0.22430\), corresponding to \(\theta_C\simeq13^\circ\).

We employ the complete six-form-factor decomposition of the
semileptonic baryon current \citep{Cabibbo:2003cu},
\begin{align}
    \Gamma^\mu(q)
    ={}&
    \gamma^\mu
    \left[
        f_1(q^2)
        -
        g_1(q^2)\gamma_5
    \right]
    \nonumber\\
    &+
    \frac{i\sigma^{\mu\nu}q_\nu}{M_1+M_2}
    \left[
        f_2(q^2)
        -
        g_2(q^2)\gamma_5
    \right]
    \nonumber\\
    &+
    \frac{q^\mu}{M_1}
    \left[
        f_3(q^2)
        -
        g_3(q^2)\gamma_5
    \right],
    \label{eq:vertex}
\end{align}
where
\begin{equation}
    \sigma^{\mu\nu}
    =
    \frac{i}{2}
    \left[
        \gamma^\mu,\gamma^\nu
    \right].
\end{equation}
Here $M_1$ and $M_2$ are the vacuum masses of the initial and final
baryons. They enter only through the normalization convention of the
transition form factors. The quasiparticle spinors and the
phase-space kinematics instead involve the corresponding effective
masses $M_1^*$ and $M_2^*$.

The functions $f_1$, $f_2$, and $f_3$ are the vector,
weak-magnetism, and induced-scalar form factors, respectively,
whereas $g_1$, $g_2$, and $g_3$ denote the axial-vector,
weak-electricity, and induced-pseudoscalar form factors. 

The different mass factors in Equation~\eqref{eq:vertex} are part of the form-factor convention adopted here. The weak-magnetism and weak-electricity form factors \(f_2\) and \(g_2\) are normalized by \(M_1+M_2\), whereas the induced-scalar and induced-pseudoscalar form factors \(f_3\) and \(g_3\) are normalized by \(M_1\). Form-factor inputs expressed in alternative conventions are converted to this convention before use.

No truncation of Equation~\eqref{eq:vertex} is made in the general
derivation. The complete spin trace retains all quadratic and
interference terms among the six form factors. The full
phase-space expression, invariant decomposition, angular reduction,
and thermal integration are given in Appendix~\ref{app:trace}.

For strongly degenerate baryons and charged leptons,
$T\ll\mu_i$, the reaction is restricted to narrow regions around
the corresponding Fermi surfaces. To leading order in $T/\mu_i$,
all smooth quantities in the phase-space integrand can therefore be
evaluated at the Fermi surfaces, while the energy dependence of the
occupation factors and of the emitted antineutrino is retained.
As argued by \citet{Leinson:2002bv} and discussed in
Section~\ref{sec:q_vs_K}, the momentum-dependent form factors are
evaluated at the kinetic Fermi-surface transfer $q_F^2$ of
Equation~\eqref{eq:q2_fermi_surface}.

After the angular and thermal integrations, the direct-Urca
emissivity can be written as
\begin{equation}
    Q_{12\ell}
    \simeq
    \frac{457\pi}{10080}
    G_F^2 C_{12}^2
    T^6
    \Theta_{12\ell}
    \mathcal{A}_{12\ell},
    \label{eq:compact_Q}
\end{equation}
where $\mathcal{A}_{12\ell}$ is the relativistic coefficient
containing the complete quadratic and interference dependence on
the six form factors.

Its explicit quadratic decomposition is given in
Appendix~\ref{app:invariant_decomposition}. The characteristic
$T^6$ dependence of direct-Urca emission is therefore unchanged by
retaining the complete relativistic weak current.

The factor $\Theta_{12\ell}$ enforces the kinematic threshold,
\begin{align}
    \Theta_{12\ell}
    ={}&
    \Theta
    \left(
        k_{F2}+k_{F\ell}-k_{F1}
    \right)
    \nonumber\\
    &\times
    \Theta
    \left(
        k_{F1}+k_{F\ell}-k_{F2}
    \right)
    \nonumber\\
    &\times
    \Theta
    \left(
        k_{F1}+k_{F2}-k_{F\ell}
    \right).
    \label{eq:triangle_factor}
\end{align}
Thus, $\Theta_{12\ell}=1$ only when the three Fermi momenta can form
a triangle.

As a consistency check, Equation~\eqref{eq:compact_Q} reduces to the
standard relativistic nucleonic direct-Urca result of \citet{Leinson:2002bv} in the equal-baryon-mass limit with $f_3=g_2=0$.

\subsection{Redshifted Neutrino Luminosity}
\label{sec:luminosity}

To determine how the density-dependent modifications of the local
direct-Urca emissivities propagate to the stellar scale, we calculate
the total neutrino luminosity measured by an observer at infinity.

For a static, spherically symmetric star, the metric is
\begin{equation}
    ds^2
    =
    e^{2\Phi(r)}c^2dt^2
    -
    \frac{dr^2}
    {1-2Gm(r)/(rc^2)}
    -
    r^2d\Omega^2,
    \label{eq:stellar_metric}
\end{equation}
where $m(r)$ is the gravitational mass enclosed within radius $r$
and $\Phi(r)$ is the gravitational metric potential obtained from
the TOV solution.

The total local direct-Urca emissivity is
\begin{equation}
    Q_{\rm tot}(r,T)
    =
    \sum_{B_1B_2}
    \sum_{\ell=e,\mu}
    Q_{12\ell}(r,T),
    \label{eq:total_local_emissivity}
\end{equation}
where only populated and kinematically open channels contribute.
The quantity $Q_{12\ell}$ includes both beta decay and inverse
lepton capture in beta equilibrium.

The proper-volume element of a spherical shell is
\begin{equation}
    dV
    =
    \frac{
        4\pi r^2dr
    }{
        \sqrt{
            1-2Gm(r)/(rc^2)
        }
    }.
    \label{eq:proper_volume}
\end{equation}
Both the energy of an emitted neutrino and the local rate of emission
are gravitationally redshifted. The neutrino luminosity measured at
infinity is therefore
\begin{equation}
    L_\nu^\infty
    =
    4\pi
    \int_0^R
    \frac{
        r^2
        e^{2\Phi(r)}
        Q_{\rm tot}(r,T)
    }{
        \sqrt{
            1-2Gm(r)/(rc^2)
        }
    }
    \,dr,
    \label{eq:redshifted_luminosity}
\end{equation}
as in the standard general-relativistic treatment of neutron-star
cooling
\citep{Yakovlev:2000jp,Potekhin:2015qsa}.

After the stellar interior becomes thermally relaxed, it is
approximately isothermal in the redshifted temperature,
\begin{equation}
    T^\infty
    =
    T(r)e^{\Phi(r)}
    =
    {\rm const.}
    \label{eq:Tolman_temperature}
\end{equation}
We define
\begin{equation}
    T_9(r)
    \equiv
    \frac{T(r)}
    {10^9\,{\rm K}},
    \qquad
    T_9^\infty
    \equiv
    \frac{T^\infty}
    {10^9\,{\rm K}}.
    \label{eq:T9_definition}
\end{equation}

For the strongly degenerate direct-Urca processes considered here,
the total local emissivity may then be written as
\begin{equation}
    Q_{\rm tot}(r,T)
    =
    \widetilde{Q}_{\rm tot}
    \left[
        n_B(r)
    \right]
    T_9^6(r),
    \label{eq:emissivity_T9}
\end{equation}
where $\widetilde{Q}_{\rm tot}$ denotes the total emissivity evaluated
at $T=10^9\,{\rm K}$. Using
\begin{equation}
    T_9(r)
    =
    T_9^\infty
    e^{-\Phi(r)},
    \label{eq:T9_redshift}
\end{equation}
Equation~\eqref{eq:redshifted_luminosity} becomes
\begin{equation}
    \frac{
        L_\nu^\infty
    }{
        (T_9^\infty)^6
    }
    =
    4\pi
    \int_0^R
    \frac{
        r^2
        e^{-4\Phi(r)}
        \widetilde{Q}_{\rm tot}
        \left[
            n_B(r)
        \right]
    }{
        \sqrt{
            1-2Gm(r)/(rc^2)
        }
    }
    \,dr.
    \label{eq:redshifted_luminosity_T9}
\end{equation}
This is the quantity used for the stellar-sequence calculation in
Section~\ref{sec:results}.

The present calculation evaluates the instantaneous neutrino
luminosity at fixed redshifted internal temperature. We do not solve
the time-dependent thermal-evolution equations and therefore do not
construct cooling curves. The luminosity calculation is instead
intended to quantify how uncertainty in the baryonic weak vertex
propagates from the local direct-Urca rates to the integrated
neutrino power of the star.

\section{Vacuum Form-Factor Inputs and Channel Construction}
\label{sec:form_factor_inputs}

The six form factors entering Equation~\eqref{eq:vertex} describe the
weak structure of the participating baryons. For transitions between
asymptotic spin-$1/2$ baryons in vacuum, Lorentz invariance implies
that they depend only on the invariant momentum transfer,
\begin{equation}
    X_i^{\rm vac}
    =
    X_i^{\rm vac}(q^2),
    \qquad
    X_i
    \in
    \left\{
        f_1,f_2,f_3,g_1,g_2,g_3
    \right\}.
    \label{eq:vacuum_ff_definition}
\end{equation}
These are the quantities determined in phenomenological analyses and
lattice-QCD calculations of semileptonic baryon matrix elements
\citep{Cabibbo:2003cu,Sasaki:2008ha,Bacchio:2025auj}.

As discussed in Section~\ref{sec:q_vs_K}, the complete weak vertex
in dense matter is more general. The medium introduces the additional
four-vector $u^\mu$, and the coefficients of the allowed Lorentz
structures may therefore depend on the thermodynamic state of the
matter in addition to the kinematic invariants. The vacuum functions
$X_i^{\rm vac}(q^2)$ used below should consequently be regarded as
inputs to the RMF quasiparticle one-body current, rather than as a
first-principles determination of the complete finite-density weak
vertex.

Possible intrinsic density-dependent modifications of the transition
form factors are not included in the present calculation. Such
effects constitute an additional systematic uncertainty that is
distinct from the statistical uncertainties of the vacuum
lattice-QCD inputs and cannot presently be quantified for the
hyperonic transitions and densities considered here.

\subsection{Embedding of the Vacuum Form Factors}
\label{sec:ff_embedding}

Conventional hyperonic direct-Urca calculations commonly employ
zero-momentum transition couplings related through flavor SU(3)
\citep{Prakash:1992zng,Cabibbo:2003cu}. We use this construction as
our static reference.

For an isolated vacuum decay,
\begin{equation}
    B_1
    \rightarrow
    B_2+\ell+\bar{\nu}_\ell,
\end{equation}
the physical timelike interval, neglecting the neutrino mass, is
\begin{equation}
    m_\ell^2
    \leq
    q^2
    \leq
    \left(M_1-M_2\right)^2.
    \label{eq:vacuum_physical_region}
\end{equation}
This interval should not be identified with the kinetic transfer
sampled by baryonic quasiparticles in dense RMF matter. As derived in Section~\ref{sec:q_vs_K}, the Fermi-surface transfer entering the adopted quasiparticle current is
\begin{equation}
    q_F^2
    =
    m_\ell^2
    -
    2\mu_\ell\Delta U
    +
    \Delta U^2,
    \label{eq:qF2_embedding}
\end{equation}
where $\Delta U=U_1-U_2$ is the difference between the vector
single-particle potentials of the initial and final baryons.

For transitions between different baryonic species, $q_F^2$ is
therefore not constrained to the physical kinematic interval of the
corresponding isolated decay. In particular, some of the hyperonic
direct-Urca reactions considered below probe positive timelike
transfers substantially larger than those accessible in their vacuum
decays.

Evaluating a vacuum parametrization $X_i^{\rm vac}(q_F^2)$ at such
kinematics constitutes a continuation of the vacuum matrix element,
rather than a determination of the true finite-density form factor.
We employ this continuation deliberately to estimate how sensitive
the direct-Urca rate can be to the momentum dependence of the weak
baryonic vertex. The resulting calculation should therefore not be
interpreted as a first-principles prediction of the form factors in
dense matter.

To distinguish changes in the zero-momentum input from those
associated with the momentum-transfer dependence, we consider three
successive prescriptions,
\begin{align}
    X_i^{\rm SU(3)}
    &=
    X_i^{\rm SU(3)}(0),
    \label{eq:su3_zero_prescription}
    \\
    X_i^{(0)}
    &=
    X_i^{\rm lat/hyb}(0),
    \label{eq:zero_ff_prescription}
    \\
    X_i^{(\rm kin)}
    &=
    X_i^{\rm lat/hyb}(q_F^2).
    \label{eq:kinetic_ff_prescription}
\end{align}
The first prescription defines the conventional static SU(3)
reference. The comparison between
$X_i^{\rm SU(3)}$ and $X_i^{(0)}$ isolates changes in the
zero-momentum transition amplitudes arising from the adopted
lattice-informed input. The comparison between $X_i^{(0)}$ and
$X_i^{(\rm kin)}$ then isolates the additional sensitivity to the
assumed vacuum momentum dependence evaluated at the dense-matter
kinetic transfer.

The uncertainty associated with the third prescription is not fully
described by the statistical covariance of the lattice calculations.
Continuation to kinematics not directly constrained in vacuum, the
hybrid construction used for channels without direct lattice data,
and genuine finite-density modifications of the weak vertex introduce
additional systematic uncertainties that are not assigned a
quantitative error in the present work.

\subsection{Static SU(3) Reference}
\label{sec:ff_su3}

The static reference is constructed from the flavor-SU(3)
decomposition
\begin{align}
    f_1^{B_1B_2}(0)
    &=
    a_{12}F_1^V(0)
    +
    b_{12}D_1^V(0),
    \label{eq:su3_f1_static}
    \\
    f_2^{B_1B_2}(0)
    &=
    a_{12}F_2^V(0)
    +
    b_{12}D_2^V(0),
    \label{eq:su3_f2_static}
    \\
    g_1^{B_1B_2}(0)
    &=
    a_{12}F_A(0)
    +
    b_{12}D_A(0).
    \label{eq:su3_g1_static}
\end{align}
The coefficients $a_{12}$ and $b_{12}$ are listed in
Table~\ref{tab:su3_coefficients}. They follow the standard
flavor-SU(3) phase convention used for hyperon semileptonic
transitions \citep{Singh:2006xp}.

\begin{deluxetable}{ccc}
\tablecaption{
Flavor-SU(3) coefficients used for the transition form factors.
\label{tab:su3_coefficients}
}
\tablehead{
    \colhead{Transition} &
    \colhead{$a_{12}$} &
    \colhead{$b_{12}$}
}
\startdata
$n\to p$
&
$1$
&
$1$
\\
$\Lambda\to p$
&
$-\sqrt{3/2}$
&
$-\sqrt{1/6}$
\\
$\Sigma^-\to n$
&
$-1$
&
$1$
\\
$\Xi^-\to\Lambda$
&
$\sqrt{3/2}$
&
$-\sqrt{1/6}$
\\
$\Xi^-\to\Xi^0$
&
$1$
&
$-1$
\\
$\Xi^0\to\Sigma^+$
&
$1$
&
$1$
\enddata
\end{deluxetable}

The reduced vector charges are obtained through the
conserved-vector-current construction from the proton and neutron
electromagnetic form factors. The axial charges are fixed using
\begin{equation}
    F_A(0)+D_A(0)=1.2670,
    \qquad
    F_A(0)-D_A(0)=-0.341,
    \label{eq:su3_axial_charges}
\end{equation}
as determined from hyperon semileptonic decays
\citep{Cabibbo:2003cu,Singh:2006xp}.

In the static reference, the second-class form factors are set to
\begin{equation}
    f_3^{B_1B_2}(0)
    =
    g_2^{B_1B_2}(0)
    =
    0.
    \label{eq:static_second_class_zero}
\end{equation}
This follows from the assumptions of exact flavor symmetry and the
absence of second-class currents, rather than from a kinematic
equality of the initial and final baryon masses
\citep{Cabibbo:2003cu,Sasaki:2008ha}.

The induced-pseudoscalar form factor is retained using the
pseudoscalar-pole prescription of
Equation~\eqref{eq:hybrid_g3_pcac}, evaluated at $q^2=0$. This set
defines the static SU(3) reference used throughout the numerical
comparisons.

\subsection{Direct Lattice-QCD Determinations}
\label{sec:ff_direct_lqcd}

For the nucleonic transition
$n\rightarrow p+\ell+\bar{\nu}_\ell$, we use continuum-limit
lattice-QCD determinations of the proton and isovector
electromagnetic form factors and of the isovector axial and
induced-pseudoscalar form factors
\citep{Alexandrou:2025vto,Alexandrou:2023qbg}.

In the isospin-symmetric limit,
\begin{align}
    f_1^{np}(q^2)
    &=
    F_1^{p-n}(q^2),
    \label{eq:np_f1}
    \\
    f_2^{np}(q^2)
    &=
    F_2^{p-n}(q^2),
    \label{eq:np_f2}
    \\
    g_1^{np}(q^2)
    &=
    G_A(q^2),
    \label{eq:np_g1}
    \\
    g_3^{np}(q^2)
    &=
    \frac{M_n}{M_n+M_p}
    G_P(q^2).
    \label{eq:np_g3}
\end{align}
This is the conversion implemented in the numerical calculation. The
second-class nucleon form factors vanish in the isospin-symmetric
limit,
\begin{equation}
    f_3^{np}(q^2)
    =
    g_2^{np}(q^2)
    =
    0.
    \label{eq:nucleon_second_class}
\end{equation}

For the strangeness-changing transition
$\Lambda\rightarrow p+\ell+\bar{\nu}_\ell$, we use the direct
lattice-QCD determination of all six form factors from
\citet{Bacchio:2025auj}. That calculation was performed on one
physical-point $N_f=2+1+1$ ensemble with
$a\simeq0.08\,{\rm fm}$ and $L\simeq5\,{\rm fm}$. It provides
correlated determinations of the six form factors
\begin{equation}
    \left\{
        f_1,f_2,f_3,
        g_1,g_2,g_3
    \right\}
\end{equation}
and a parametrization of their vacuum momentum dependence. A
continuum or finite-volume extrapolation is not yet available.

For the $\Lambda\rightarrow p$ channel, unlike the hybrid channels introduced below,
all six structures of the weak current can be retained using direct
transition-specific lattice information. It consequently provides a benchmark for examining the relative importance of the leading and second-class form factors and their interference in the direct-Urca rate.

\subsection{Hybrid Hyperon Form Factors}
\label{sec:ff_hybrid}

A comparably complete physical-point lattice-QCD determination is not
presently available for all remaining hyperonic transitions. We
therefore construct
\begin{equation}
    \Sigma^-\to n,
    \qquad
    \Xi^-\to\Lambda,
    \qquad
    \Xi^-\to\Xi^0,
    \qquad
    \Xi^0\to\Sigma^+
\end{equation}
using a hybrid prescription. The flavor-SU(3) coefficients determine
the transition quantum numbers and zero-momentum structure, while the
momentum dependence is inherited from continuum-limit nucleon
lattice-QCD form factors.

Not all of these reactions are kinematically active for the FSU2H
EOS used in the numerical results. Nevertheless, we define the
complete set because channel activation is an EOS-dependent result
determined by the particle populations and the triangle factor
$\Theta_{12\ell}$, rather than by the form-factor model. Retaining
the full set also allows the same construction to be applied to
alternative equations of state.

The corresponding reduced SU(3) functions are
\begin{align}
    F_i^V(q^2)
    &=
    f_i^p(q^2)
    +
    \frac{1}{2}f_i^n(q^2),
    \nonumber\\
    D_i^V(q^2)
    &=
    -\frac{3}{2}f_i^n(q^2),
    \qquad
    i=1,2,
    \label{eq:hybrid_vector_reduced}
\end{align}
as follows from the conserved-vector-current construction
\citep{Cabibbo:2003cu,Singh:2006xp}. The transition form factors are
then
\begin{align}
    f_1^{B_1B_2}(q^2)
    &=
    a_{12}F_1^V(q^2)
    +
    b_{12}D_1^V(q^2),
    \label{eq:hybrid_f1}
    \\
    f_2^{B_1B_2}(q^2)
    &=
    a_{12}F_2^V(q^2)
    +
    b_{12}D_2^V(q^2).
    \label{eq:hybrid_f2}
\end{align}

For the axial sector, separate momentum-dependent determinations of
the reduced functions $F_A$ and $D_A$ are not available for the
hyperonic transitions considered here. We therefore assume, following
the standard SU(3)-based construction
\citep{Singh:2006xp}, that the two reduced functions share the
momentum dependence of the nucleon isovector axial form factor
$G_A(q^2)$. Defining
\begin{align}
    F_A(q^2)
    &=
    x_A G_A(q^2),
    \nonumber\\
    D_A(q^2)
    &=
    \left(
        1-x_A
    \right)
    G_A(q^2),
    \nonumber\\
    x_A
    &=
    \frac{
        F_A(0)
    }{
        F_A(0)+D_A(0)
    },
    \label{eq:hybrid_axial_reduced}
\end{align}
we use
\begin{equation}
    g_1^{B_1B_2}(q^2)
    =
    a_{12}F_A(q^2)
    +
    b_{12}D_A(q^2).
    \label{eq:hybrid_g1}
\end{equation}

The hybrid construction is not a lattice-QCD determination of the
hyperonic transition form factors. Lattice QCD supplies the nucleon
momentum dependence, while its transfer to the hyperonic channels
through the SU(3) decomposition is a model assumption. In
particular, channel-dependent SU(3)-breaking effects in the shapes of
the form factors are not known and are not included here. The hybrid
prescription is therefore used to estimate the characteristic
momentum-transfer sensitivity of channels for which direct
transition-specific information is unavailable.

In the absence of channel-specific information, the induced-scalar
and weak-electricity form factors are set to
\begin{equation}
    f_3^{B_1B_2}(q^2)
    =
    g_2^{B_1B_2}(q^2)
    =
    0.
    \label{eq:hybrid_second_class_zero}
\end{equation}
This is an explicit model assumption and is not a consequence of the
dense-matter kinematics. Earlier lattice calculations demonstrate
that second-class form factors can become nonzero once flavor SU(3)
is broken \citep{Sasaki:2008ha}. The direct
$\Lambda\rightarrow p$ calculation described above is used to
examine separately how important such terms can be when they are
retained explicitly.

The induced-pseudoscalar form factor is modeled through
pseudoscalar-pole dominance,
\begin{equation}
    g_3^{B_1B_2}(q^2)
    =
    \frac{
        M_1\left(M_1+M_2\right)
    }{
        m_P^2-q^2
    }
    g_1^{B_1B_2}(q^2),
    \label{eq:hybrid_g3_pcac}
\end{equation}
where
\begin{equation}
    m_P
    =
    \begin{cases}
        m_\pi, & \Delta S=0,\\
        m_K,   & |\Delta S|=1.
    \end{cases}
    \label{eq:hybrid_pole_mass}
\end{equation}
The sign of $g_3$ is therefore inherited from that of $g_1$ in the
common phase convention. Equation~\eqref{eq:hybrid_g3_pcac} is a
model input based on PCAC and pole dominance, rather than a direct
lattice determination
\citep{Cabibbo:2003cu,Singh:2006xp,Sasaki:2008ha}.

This distinction becomes especially important in the present
application because the kinetic transfers sampled by some hyperonic
direct-Urca channels can approach a region in which the pole
denominator varies rapidly. Predictions involving $g_3$ at such
kinematics should therefore be regarded as more model dependent than
the corresponding leading vector and axial contributions.

Where lattice information on the zero-momentum vector form factor is
available, we use it to anchor the normalization of the corresponding
hybrid form factor while retaining the nucleon-lattice-QCD momentum
dependence. Specifically,
\begin{align}
    f_1^{\Sigma^- n}(0)
    &=
    -0.9571(60),
    \nonumber\\
    f_1^{\Xi^0\Sigma^+}(0)
    &=
    0.9755(39),
    \label{eq:hybrid_f1_anchors}
\end{align}
from the lattice-informed physical-point analysis of
\citet{Sasaki:2017jue}.

The resulting hierarchy of form-factor inputs is therefore explicit.
The static SU(3) prescription provides the conventional
zero-momentum reference. The direct lattice-QCD calculations provide
transition-specific vacuum information for $n\rightarrow p$ and
$\Lambda\rightarrow p$, and the hybrid prescription extends the
known nucleon momentum dependence to the remaining hyperonic
channels through the SU(3) structure. Evaluating the latter two
classes at $q_F^2$ then provides the momentum-transfer sensitivity
studied in Section~\ref{sec:results}. No claim is made that this
procedure determines the intrinsic finite-density modification of
the baryonic weak vertex.

\section{Results}
\label{sec:results}

\begin{figure*}
\centering
\includegraphics[width=0.85\linewidth]
{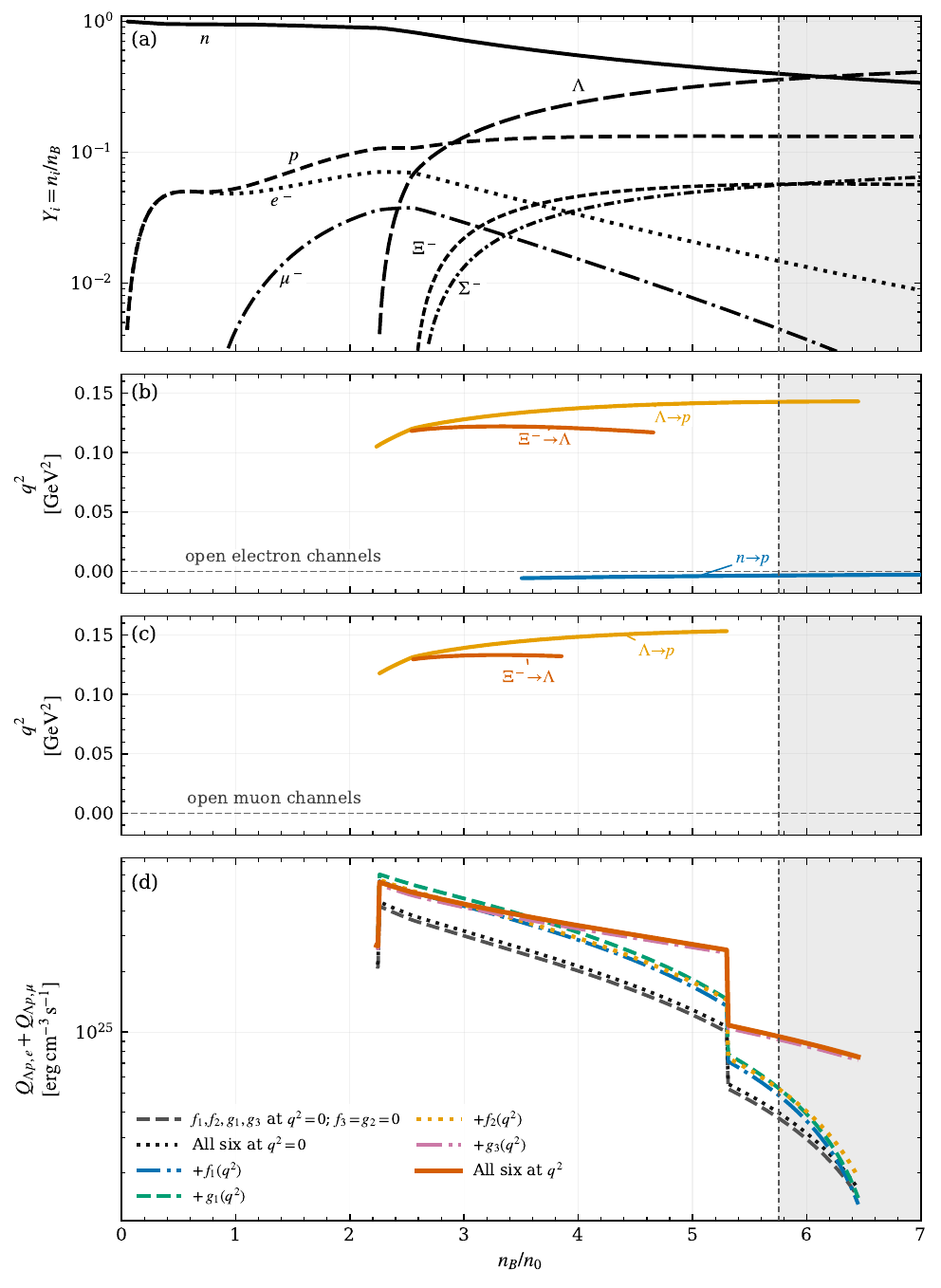}
\caption{
Density dependence of the FSU2H composition, direct-Urca kinematics,
and form-factor sensitivity of the $\Lambda\rightarrow p$ channel.
Panel (a) shows the particle fractions $Y_i=n_i/n_B$.
Panels (b) and (c) show the kinetic invariant transfer
$q_F^2=(P_1-P_2)^2$ sampled by the kinematically open electron and
muon direct-Urca branches, respectively. Panel (d) shows
$Q_{\Lambda p,e}+Q_{\Lambda p,\mu}$ while progressively introducing
the $q_F^2$ dependence of the six directly calculated
$\Lambda\rightarrow p$ form factors.
The vertical dashed line denotes the central density of the
maximum-mass stable FSU2H star. The shaded region is not reached by
stable stellar configurations.
}
\label{fig:density_kinematics}
\end{figure*}

\begin{figure}
\centering
\includegraphics[width=\linewidth]
{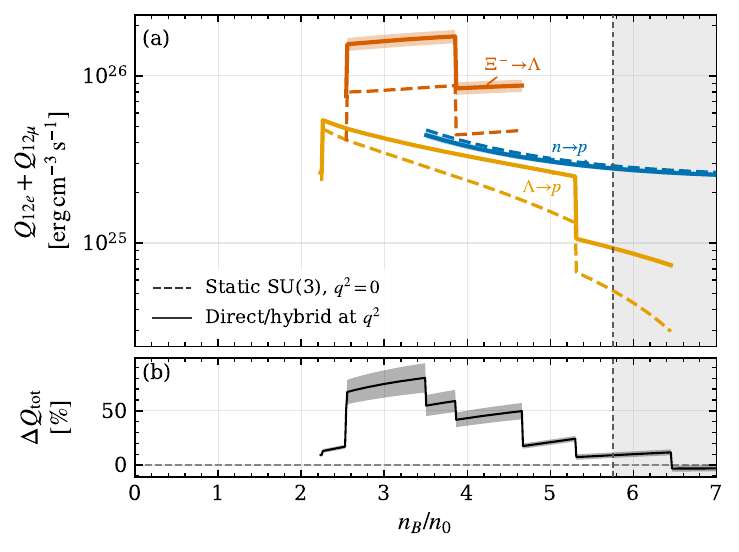}
\caption{
Direct-Urca emissivities in FSU2H matter at $T_9=1$.
Panel (a) shows the electron-plus-muon emissivity of each
kinematically active transition. Dashed curves denote the static
flavor-SU(3) reference evaluated at $q^2=0$, while solid curves use
the direct or hybrid lattice-informed form factors evaluated at the
kinetic Fermi-surface transfer $q_F^2$.
Panel (b) shows the relative difference of the total emissivity,
summed over all open baryonic and leptonic channels,
$\Delta Q_{\rm tot}
=100\,(Q_{\rm tot}^{(\rm kin)}-Q_{\rm tot}^{\rm SU(3)})
/Q_{\rm tot}^{\rm SU(3)}$.
The shaded band represents the propagated $1\sigma$ uncertainty from
the form-factor fit covariances included in the calculation.
The vertical dashed line and shaded region have the same meaning as
in Figure~\ref{fig:density_kinematics}.
}
\label{fig:channel_emissivities}
\end{figure}

\begin{figure}
\centering
\includegraphics[width=\linewidth]
{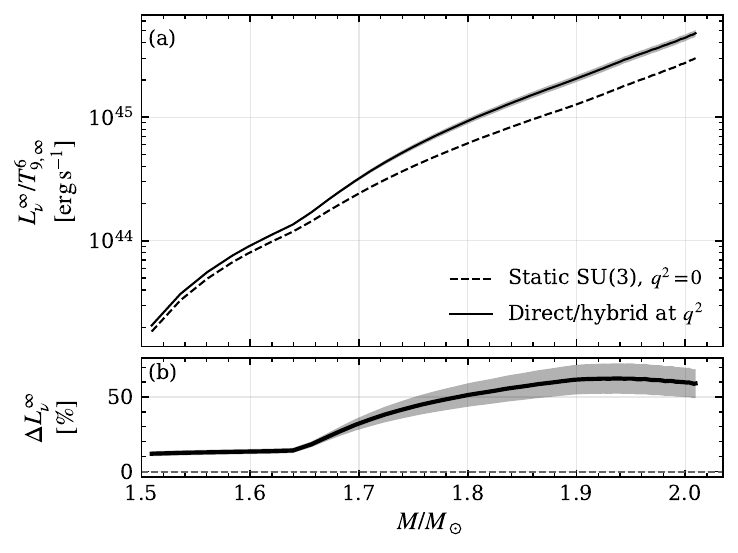}
\caption{
Redshifted neutrino luminosity along the stable FSU2H stellar
sequence. The upper panel shows
$L_\nu^\infty/T_{9,\infty}^{6}$ for the static SU(3) reference and
the lattice-informed kinetic-transfer prescription.
The lower panel shows
$\Delta L_\nu^\infty
=100\,(L_{\nu,{\rm kin}}^\infty-L_{\nu,{\rm SU(3)}}^\infty)
/L_{\nu,{\rm SU(3)}}^\infty$.
}
\label{fig:luminosity}
\end{figure}

The results presented below are evaluated at $T_9=1$. Within the
strongly degenerate approximation adopted in
Section~\ref{sec:current}, all direct-Urca emissivities scale as
$T^6$. Consequently, the relative differences between the
form-factor prescriptions are independent of temperature at this
order, provided that the EOS composition and the intrinsic form
factors are treated as temperature independent.

The numerical comparisons should be interpreted according to the
hierarchy introduced in Section~\ref{sec:form_factor_inputs}.
The static SU(3) result provides the conventional reference, while
the lattice-informed calculations test the sensitivity to changes in
the vacuum form-factor input and to its continuation to the kinetic
transfer $q_F^2$ sampled in RMF matter.

\subsection{Composition and momentum transfer}
\label{sec:results_kinematics}

Figure~\ref{fig:density_kinematics} summarizes the particle
composition and the direct-Urca kinematics of FSU2H matter.
Panel (a) shows the particle fractions. The $\Lambda$ is the first
hyperon to appear, at approximately
$n_B\simeq2.2\,n_0$, followed at higher density by negatively charged
hyperons. The vertical dashed line denotes the central density of the
maximum-mass stable FSU2H configuration, $n_c(M_{\rm max}) \simeq 5.8\,n_0$. Matter at densities to the right of this line is therefore not
sampled by stable FSU2H stars.

The presence of the required particle species is not sufficient by
itself to activate a direct-Urca reaction. The corresponding Fermi
momenta must additionally satisfy the triangle condition of
Equation~\eqref{eq:triangle_factor}. For the FSU2H composition, the
channels that become kinematically allowed over the density interval
relevant to stable stars are
\begin{equation}
    n\rightarrow p,
    \qquad
    \Lambda\rightarrow p,
    \qquad
    \Xi^-\rightarrow\Lambda.
\end{equation}
The remaining hyperonic transitions defined in
Section~\ref{sec:form_factor_inputs} remain closed over this density
range. The sequence of active reactions is therefore determined
jointly by the appearance of the relevant particles and by their
Fermi-momentum geometry.

Panels (b) and (c) of
Figure~\ref{fig:density_kinematics} show the kinetic invariant
transfer $q_F^2$ for the electron and muon branches, respectively.
A clear separation is found between the nucleonic and hyperonic
reactions. For $n\rightarrow p$, the kinetic transfer remains close
to zero throughout the density interval in which the reaction is
active. The smallness of $q_F^2$ reflects the relatively similar
vector mean fields experienced by the neutron and proton and the
correspondingly small difference between the canonical and kinetic
transfers.

The hyperonic reactions behave very differently. Both
$\Lambda\rightarrow p$ and $\Xi^-\rightarrow\Lambda$ sample
positive timelike transfers of characteristic magnitude
\begin{equation}
    q_F^2
    \simeq
    0.1\text{--}0.15~{\rm GeV}^2
\end{equation}
over substantial parts of their allowed density intervals.
The electron and muon branches probe similar scales, with smaller
differences arising from their Fermi momenta and from the densities
at which the individual branches satisfy the direct-Urca threshold.

This behavior follows directly from the RMF relation
\begin{equation}
    q_F^2
    =
    m_\ell^2
    -
    2\mu_\ell\Delta U
    +
    \Delta U^2,
\end{equation}
derived in Section~\ref{sec:q_vs_K}. The large hyperonic values are therefore caused by the difference in vector mean-field energies between the initial and final baryonic
quasiparticles.

This is the central kinematic result of the present calculation.
For the nucleonic process, the weak vertex samples the form factors
close to their zero-momentum values. Hyperonic reactions can instead
probe a substantially larger timelike momentum-transfer region.
For several of these reactions the resulting $q_F^2$ extends beyond
the physical interval of the corresponding isolated semileptonic
decay. The RMF calculation therefore determines the kinematics at
which the quasiparticle current is sampled, but existing vacuum
information does not uniquely determine the form factors at these
kinematics. The finite-$q_F^2$ calculations below quantify the
sensitivity to plausible continuations of the available vacuum input.

\subsection{Form-factor sensitivity of the
$\Lambda\rightarrow p$ channel}
\label{sec:results_lambda}

The $\Lambda\rightarrow p$ transition provides the cleanest channel
with which to investigate the structure of the weak current because
all six transition form factors are available directly from the same
lattice-QCD calculation. Panel (d) of
Figure~\ref{fig:density_kinematics} shows the total
$\Lambda\rightarrow p$ emissivity,  $ Q_{\Lambda p}
    =
    Q_{\Lambda p,e}
    +
    Q_{\Lambda p,\mu}$, while progressively introducing the momentum dependence of the
individual form factors.

The two lowest-level prescriptions in this comparison are evaluated
entirely at $q^2=0$. Comparing the four-form-factor result, for which
$f_3=g_2=0$, with the complete six-form-factor result shows that the
directly calculated second-class form factors modify the
zero-momentum emissivity only moderately. Thus, the presence of
nonzero $f_3$ and $g_2$ does not by itself generate the large
difference observed when the dense-matter kinetic transfer is used.

A substantially larger change appears when the momentum dependence
of the form factors is introduced. Allowing $f_1$ to vary with
$q_F^2$ first increases the emissivity relative to the complete
$q^2=0$ result. Including the momentum dependence of $g_1$ produces
an additional increase. The subsequent inclusion of
$f_2(q_F^2)$ gives a sizeable compensating contribution. This
demonstrates that the response cannot be understood by considering
only the dominant vector or axial form factor independently.

The inclusion of the induced-pseudoscalar form factor
$g_3(q_F^2)$ produces another important change and brings the result
close to that obtained when all six form factors are evaluated at
$q_F^2$. The remaining effect of introducing the momentum dependence
of $f_3$ and $g_2$ is comparatively small. For
$\Lambda\rightarrow p$, the finite-$q_F^2$ sensitivity therefore
arises primarily through the combined momentum dependence and
interference of $f_1$, $g_1$, $f_2$, and $g_3$, rather than from the
second-class currents alone.

The relatively large pseudoscalar contribution warrants further
discussion. In an isolated vacuum decay, the momentum entering the
baryonic vertex is identical to the momentum carried by the leptonic
system,
\begin{equation}
    q^\mu
    =
    K^\mu
    =
    k_\ell^\mu+k_{\nu_\ell}^\mu.
\end{equation}
For a massless antineutrino,
\begin{equation}
    K_\mu
    \bar u_\ell
    \gamma^\mu(1-\gamma_5)
    v_{\nu_\ell}
    =
    m_\ell
    \bar u_\ell
    (1-\gamma_5)
    v_{\nu_\ell}.
    \label{eq:g3_vacuum_suppression}
\end{equation}
The induced-pseudoscalar contribution is consequently suppressed by
the charged-lepton mass in the usual vacuum kinematics.

For the RMF quasiparticle current considered here,
\begin{equation}
    q^\mu
    =
    K^\mu-\Delta U\,u^\mu,
\end{equation}
and therefore
\begin{align}
    q_\mu
    \bar u_\ell
    \gamma^\mu(1-\gamma_5)
    v_{\nu_\ell}
    ={}&
    m_\ell
    \bar u_\ell
    (1-\gamma_5)
    v_{\nu_\ell}
    \nonumber\\
    &-
    \Delta U\,
    \bar u_\ell
    \gamma^0(1-\gamma_5)
    v_{\nu_\ell}.
    \label{eq:g3_medium_contraction}
\end{align}
The second term is not proportional to $m_\ell$. Within the adopted
$q^\mu$-based quasiparticle vertex, the induced-pseudoscalar
contribution is therefore not constrained by the same lepton-mass
suppression as in an isolated decay. This accounts for its visible
effect also in the electron branch.

For the strangeness-changing channels there is an additional
sensitivity associated with the pole-dominance form used for the
induced-pseudoscalar contribution. In the hybrid construction,
\begin{equation}
    g_3(q^2)
    =
    \frac{
        M_1(M_1+M_2)
    }{
        m_K^2-q^2
    }
    g_1(q^2).
    \label{eq:g3_result_discussion}
\end{equation}
The hyperonic values
$q_F^2\simeq0.1$--$0.15~{\rm GeV}^2$ are significantly closer to
the kaon-pole scale than $q^2=0$, making the resulting $g_3$
contribution particularly sensitive to the assumed continuation.

The sizeable pseudoscalar response should therefore not be
interpreted as a precise prediction of $g_3$ in dense matter.
Rather, it demonstrates that induced-pseudoscalar physics may become
important once the kinetic and canonical momentum transfers differ,
and that a transition-specific determination of the weak vertex at
finite density would be required to establish its magnitude.

\subsection{Channel-resolved and total direct-Urca emissivities}
\label{sec:results_emissivity}

Figure~\ref{fig:channel_emissivities} compares the conventional
static SU(3) reference with the direct or hybrid lattice-informed
prescription evaluated at the kinetic momentum transfer. Panel (a)
shows the emissivity of each kinematically active transition after
summing the electron and muon contributions, $Q_{12}
    =
    Q_{12e}
    +
    Q_{12\mu}$.
    
The discontinuous changes visible in some curves arise from the
opening or closing of individual lepton branches through the
Fermi-momentum triangle condition and are therefore kinematic rather
than discontinuities of the form factors themselves.

The $n\rightarrow p$ emissivity is only weakly modified by replacing
the static input with the direct lattice-QCD form factors evaluated
at $q_F^2$. This is consistent with
Figure~\ref{fig:density_kinematics}, where the nucleonic process
samples $q_F^2$ very close to zero. The nucleonic rate is therefore
mainly sensitive to differences in the zero-momentum normalization
of the input and only weakly to its momentum dependence.

The hyperonic channels show a much stronger dependence on the
form-factor prescription. The $\Lambda\rightarrow p$ emissivity
differs substantially from its static SU(3) reference over much of
the density interval in which the reaction is active. An even larger
difference is obtained for $\Xi^-\rightarrow\Lambda$ over part of
its allowed range. These are precisely the channels for which the
RMF kinematics generate kinetic transfers of order
$0.1~{\rm GeV}^2$ or larger.

For $\Lambda\rightarrow p$, the availability of direct
transition-specific lattice-QCD information allows the origin of
this sensitivity to be studied in some detail, as discussed above.
For $\Xi^-\rightarrow\Lambda$, in contrast, the result relies on the
hybrid construction of Section~\ref{sec:ff_hybrid}. Its numerical
magnitude is consequently more model dependent. In particular, the
transfer of nucleon momentum dependence to a hyperonic transition
through flavor SU(3) is not a first-principles determination of the
$\Xi^-\rightarrow\Lambda$ form factors at large timelike $q_F^2$.
The channel should therefore be viewed primarily as an estimate of
the possible scale of the sensitivity.

To quantify the effect on the complete direct-Urca emission, we
define
\begin{equation}
    Q_{\rm tot}
    =
    \sum_{B_1B_2}
    \left(
        Q_{12e}
        +
        Q_{12\mu}
    \right),
    \label{eq:results_Qtot}
\end{equation}
where the sum contains all kinematically open channels, and
\begin{equation}
    \Delta Q_{\rm tot}
    =
    100\,
    \frac{
        Q_{\rm tot}^{(\rm kin)}
        -
        Q_{\rm tot}^{\rm SU(3)}
    }{
        Q_{\rm tot}^{\rm SU(3)}
    }.
    \label{eq:results_deltaQ}
\end{equation}
Panel (b) of Figure~\ref{fig:channel_emissivities} shows this
quantity as a function of density.

The difference between the two prescriptions is strongly density
dependent because both the relative importance of the individual
baryonic transitions and the availability of their electron and
muon branches change with density. Following the onset of hyperonic
direct Urca, $\Delta Q_{\rm tot}$ reaches several tens of percent
and increases as the hyperonic channels become more important. In
the present implementation, the largest difference is approximately
$80\%$ around $n_B
    \simeq
    3\text{--}4\,n_0$.
    
At higher densities the difference decreases in a stepwise manner as
the active-channel composition changes. Near the central density of
the maximum-mass stable FSU2H star, it is considerably smaller, of
order $10\%$.

The value of $\Delta Q_{\rm tot}$ should not be interpreted as a
quantified correction to the physical emissivity. It measures the
difference between the static-coupling reference and the particular
vacuum-form-factor continuation adopted here. The result nevertheless
demonstrates that, once hyperonic channels sample kinetic transfers
of order $0.1~{\rm GeV}^2$, uncertainty in the weak transition
structure can become comparable to tens of percent or larger.

\subsection{Neutrino luminosity of FSU2H stars}
\label{sec:results_luminosity}

The density-dependent emissivities are finally integrated over the
stellar profiles obtained from the FSU2H EOS. The
general-relativistic luminosity integral and redshifted-isothermal
temperature prescription are given in Section~\ref{sec:luminosity}.
Since the local emissivity scales as $T^6$, we present the
redshifted luminosity in the temperature-independent form
       $ L_\nu^\infty/T_{9,\infty}^6.$

The upper panel of Figure~\ref{fig:luminosity} shows this quantity
along the stable stellar sequence as a function of gravitational
mass. The neutrino luminosity increases rapidly with stellar mass in
both form-factor prescriptions. As the central density increases,
an increasing fraction of the stellar core satisfies the thresholds
for the fast direct-Urca reactions, including the hyperonic channels.

For the comparison shown here, the lattice-informed
kinetic-transfer prescription gives a larger integrated luminosity
than the static SU(3) reference. We define the luminosity relative difference as
\begin{equation}
    \Delta L_\nu^\infty
    =
    100\,
    \frac{
        L_{\nu,{\rm kin}}^\infty
        -
        L_{\nu,{\rm SU(3)}}^\infty
    }{
        L_{\nu,{\rm SU(3)}}^\infty
    }.
    \label{eq:results_deltaL}
\end{equation}

The lower panel of Figure~\ref{fig:luminosity} shows this relative difference. For the lower-mass configurations displayed, $\Delta L_\nu^\infty$ is approximately $10$--$15\%$. It then increases with stellar mass, reaching approximately \(40\%\) around \(M\simeq1.7\!-\!1.8\,M_\odot\) and approximately \(60\%\) near \(M\simeq1.9\!-\!2.0\,M_\odot\).
A mild decrease occurs very close
to the maximum-mass configuration, although the integrated
difference remains substantial.

An important feature of Figure~\ref{fig:luminosity} is that the
stellar luminosity does not simply reflect the local emissivity
difference at the central density. In the most massive stable
configurations, $\Delta Q_{\rm tot}$ at the stellar center is
comparatively modest, whereas the integrated luminosity can still
differ by approximately $50$--$60\%$. The luminosity samples the
entire stellar density profile rather than only the central point.
Massive stars contain extended shells in the density interval in
which $\Lambda\rightarrow p$ and $\Xi^-\rightarrow\Lambda$ are
active and where the sensitivity to the form-factor prescription is
largest. These regions can therefore dominate the integrated
difference even when the highest-density central matter is less
sensitive.

As for the local emissivity, these luminosity differences should not
be interpreted as precision predictions of a finite-density
form-factor correction. They quantify how strongly the integrated
neutrino power can respond to the weak-vertex prescription when
vacuum form factors are continued to the RMF kinetic transfer.
The uncertainty of this procedure is presently not specified. The
vacuum parametrizations are extrapolated beyond directly constrained
kinematics for some channels, the remaining hyperonic transitions
use a hybrid SU(3)--nucleon construction, and genuine intrinsic
finite-density modifications of the weak vertex are not known.

As a check of the dependence on the dense-matter model, we repeated
the complete calculation using the hyperonic FSU2R parametrization
of \citet{Tolos:2016hhl}. The qualitative hierarchy is preserved.
The nucleonic $n\to p$ channel probes $q_F^2\simeq0$, whereas the
hyperonic $\Lambda\to p$ and $\Xi^-\to\Lambda$ channels sample
substantially larger timelike momentum transfers and remain more
sensitive to the momentum-dependent weak vertex. Quantitatively,
however, both the local emissivity correction and the integrated
luminosity correction change appreciably with the EOS. This is
expected because the particle composition, mean fields, reaction
thresholds, and stellar density profiles are EOS dependent. The
FSU2R results are shown in Appendix~\ref{app:eos_dependence}, but a systematic study of EOS and hyperon-interaction uncertainties is
beyond the scope of the present work.

Baryonic pairing is also not included in the present comparison.
The absolute emissivities and luminosities should therefore be
regarded as an unpaired baseline designed to isolate the sensitivity
to baryonic weak-current structure. A quantitative cooling
prediction would additionally require pairing gaps, thermal
evolution, and a better constrained finite-density weak vertex.

\section{Conclusions}
\label{sec:conclusions}

We investigated the sensitivity of direct-Urca neutrino emission in
hyperonic neutron-star matter to the structure and momentum dependence
of the baryonic charged weak current. We derived the relativistic
emissivity for general spin-$1/2$ baryon transitions retaining all six
form factors and their interference terms, and applied it to
beta-equilibrated FSU2H matter. The calculation distinguishes the
canonical momentum transferred to the lepton pair from the kinetic
momentum transfer $q^\mu$ entering the adopted RMF quasiparticle
current.

The main kinematic result is that the momentum transfer is strongly
channel dependent. For $n\rightarrow p$, $q_F^2$ remains close to
zero, whereas the hyperonic $\Lambda\rightarrow p$ and
$\Xi^-\rightarrow\Lambda$ channels can probe positive timelike values
$q_F^2\simeq0.1$--$0.15~{\rm GeV}^2$. Hyperonic direct-Urca rates can
therefore be considerably more sensitive to the momentum dependence
of the weak form factors than the nucleonic process. For
$\Lambda\rightarrow p$, where all six form factors are available
directly from lattice QCD, the dominant finite-$q^2$ sensitivity
arises from the combined momentum dependence and interference of
$f_1$, $g_1$, $f_2$, and $g_3$, while the second-class form factors
$f_3$ and $g_2$ produce a comparatively smaller effect.

Within the vacuum-form-factor continuation studied here, the total
local direct-Urca emissivity differs from the static SU(3) reference
by as much as approximately $80\%$ around
$n_B\simeq3$--$4\,n_0$, while the difference decreases to about
$10\%$ near the central density of the maximum-mass stable
configuration. 
After integration over the stellar interior, the redshifted neutrino luminosity differs by approximately \(10\!-\!15\%\) for the lower-mass configurations shown, about \(40\%\) around \(1.7\!-\!1.8\,M_\odot\), and up to approximately \(60\%\) near \(1.9\!-\!2.0\,M_\odot\). The stellar effect does not simply track the
central-density correction because the luminosity samples extended
density regions in which the hyperonic channels are active.

Repeating the analysis with the hyperonic FSU2R EOS preserves the
qualitative enhancement of the hyperonic momentum-transfer
sensitivity but changes its quantitative magnitude, demonstrating
that the numerical emissivity and luminosity modifications are EOS
dependent.

These differences should not be interpreted as first-principles
predictions of the finite-density weak vertex. The lattice-QCD inputs
describe vacuum baryon matrix elements, and for some hyperonic
channels their evaluation at the RMF kinetic transfer requires a
continuation beyond the kinematic region directly constrained in
vacuum. The hybrid hyperon construction and the induced-pseudoscalar
input introduce additional model dependence, while genuine
density-dependent modifications of the weak vertex remain unknown.
Our results therefore provide an estimate of the possible scale of
this presently unquantified weak-vertex uncertainty and motivate
transition-specific calculations of baryonic form factors under
neutron-star conditions.

\section*{Acknowledgements}
I acknowledge financial support from \texttt{IMAGE-N Excellence hub/0524/0459}, co-financed by the European Regional Development Fund and the Republic of Cyprus through the Research and Innovation Foundation within the framework of the Cohesion Policy Programme “THALIA 2021-2027”, and from the projects HyperON (VISION ERC - PATH 2/0524/0001) and Baryon8 (POST-DOC/0524/0001), co-financed by the European Regional Development Fund and the Republic of Cyprus through the Research and Innovation Foundation. I would like to express my gratitude to Simone Bacchio for valuable comments, suggestions, and discussions on this paper. Laslty, I would like to thank
Kyriakos Kokkantas, Manolis Enytro, Johny Ombikou and Kostas Psoma for our insightful conversations.

\section*{Software Availability}

The numerical analysis and computational notebooks associated with this work
are available on GitHub at
\url{https://github.com/akonstantinu/Neutrino-Emission-with-LQCD-FFs}
and are archived on Zenodo at
\dataset[10.5281/zenodo.23196444  ]
{https://doi.org/10.5281/zenodo.23196444  }.
\appendix
\section{Matrix Element and Degenerate Reduction}
\label{app:trace}

This Appendix gives the complete spin-summed matrix element for the
six-form-factor current of Equation~\eqref{eq:vertex} and derives
the strongly degenerate form used in
Equation~\eqref{eq:compact_Q}. We also make explicit how canonical
energy--momentum conservation is combined with kinetic
quasiparticle momenta in the RMF formulation.

\subsection{Phase-space expression}
\label{app:phase_space}

For baryonic quasiparticles we define
\begin{equation}
    d\Pi_i
    =
    \frac{
        d^3p_i
    }{
        (2\pi)^3\,2\epsilon_i
    },
    \qquad
    i=1,2,
    \label{eq:app_baryon_phase_space}
\end{equation}
while for the charged lepton and antineutrino
\begin{equation}
    d\Pi_a
    =
    \frac{
        d^3k_a
    }{
        (2\pi)^3\,2\omega_a
    },
    \qquad
    a=\ell,{\nu_\ell}.
    \label{eq:app_lepton_phase_space}
\end{equation}

The total neutrino and antineutrino energy emitted per unit volume
and time is
\citep{Leinson:2002bv}
\begin{align}
    Q_{12\ell}
    ={}&
    2
    \int
    d\Pi_1
    d\Pi_2
    d\Pi_\ell
    d\Pi_{\nu_\ell}
    \,
    (2\pi)^4
    \delta^{(4)}
    \left(
        \mathcal{P}_1
        -
        \mathcal{P}_2
        -
        k_\ell
        -
        k_{\nu_\ell}
    \right)
    \omega_{\nu_\ell}
    \sum_{\rm spins}
    \left|
        \mathcal{M}_{12\ell}
    \right|^2
    \,
    \mathfrak{f}_1(\epsilon_1)
    \left(
        1-\mathfrak{f}_2(\epsilon_2)
    \right)
    \left(
        1-\mathfrak{f}_\ell(\omega_\ell)
    \right).
    \label{eq:app_full_Q}
\end{align}
Here the factor of two accounts for the equal contributions of
beta decay and inverse lepton capture in beta equilibrium. $\mathfrak{f}_i$ denote the Fermi--Dirac distribution functions, which are explicitly defined as

\begin{equation}
    \mathfrak{f}_1(\epsilon_1)=\frac{1}{e^{(\epsilon_1+U_1-\mu_1)/T}+1}, \qquad \mathfrak{f}_2(\epsilon_2)=\frac{1}{e^{(\epsilon_2+U_2-\mu_2)/T}+1}, \qquad \mathfrak{f}_\ell(\omega_\ell)=\frac{1}{e^{(\omega_\ell-\mu_\ell)/T}+1}.
    \label{eq:Fermi-Dirac}
\end{equation}

The delta function contains the canonical baryon momenta and
therefore enforces
\begin{equation}
    \mathcal{P}_1^\mu-\mathcal{P}_2^\mu
    =
    k_\ell^\mu+k_{\nu_\ell}^\mu
    =
    K^\mu.
    \label{eq:app_K_conservation}
\end{equation}
The baryonic spinors, in contrast, satisfy
\begin{equation}
    \left(
        \gamma_\mu P^\mu_i-M_i^*
    \right)
    u_i(P_i)
    =
    0,
\end{equation}
and their spin sums are
\begin{equation}
    \sum_s
    u_i(P_i,s)
    \bar{u}_i(P_i,s)
    =
    \gamma_\mu P^\mu_i+M_i^*.
    \label{eq:app_spin_sum}
\end{equation}
The kinetic momentum appearing in the one-body vertex is
\begin{equation}
    q^\mu
    =
    P_1^\mu-P_2^\mu
    =
    K^\mu-\Delta U\,u^\mu.
    \label{eq:app_qK}
\end{equation}

\subsection{Leptonic and hadronic tensors}
\label{app:tensors}

After summing over external spin states,
\begin{equation}
    \sum_{\rm spins}
    \left|
        \mathcal M_{12\ell}
    \right|^2
    =
    \frac{
        G_F^2 C_{12}^2
    }{2}
    L_{\mu\nu}
    H^{\mu\nu},
    \label{eq:app_spin_sum_amplitude}
\end{equation}
where 
    $L_{\alpha\beta}
    =
    {\rm Tr}
    \left[
        \left(
            \gamma_\mu k^\mu_\ell+m_\ell
        \right)
        \gamma_\alpha
         (1-\gamma_5)
        \gamma_\mu k^\mu_{\nu_\ell}
        \gamma_\beta
         (1-\gamma_5)
    \right]$
and $H^{\alpha\beta}
    =
    {\rm Tr}
    \left[
        \left(
            \gamma_\mu P^\mu_2+M_2^*
        \right)
        \Gamma^\alpha(q)
        \left(
            \gamma_\mu P^\mu_1+M_1^*
        \right)
        \overline{\Gamma}^{\,\beta}(q)
    \right]$, with $\overline{\Gamma}^{\,\beta}
    =
    \gamma^0
    \Gamma^{\beta\dagger}
    \gamma^0$.

\subsection{Invariant decomposition}
\label{app:invariant_decomposition}

Before introducing the strongly degenerate approximation, we define
the six Lorentz invariants
\begin{align}
    S
    &\equiv
    P_1\cdot P_2,
    &
    C
    &\equiv
    k_{\nu_\ell}\cdot k_\ell,
    \nonumber\\
    A_1
    &\equiv
    k_{\nu_\ell}\cdot P_1,
    &
    A_2
    &\equiv
    k_{\nu_\ell}\cdot P_2,
    \nonumber\\
    B_1
    &\equiv
    k_\ell\cdot P_1,
    &
    B_2
    &\equiv
    k_\ell\cdot P_2.
    \label{eq:app_invariants}
\end{align}
The kinetic baryon momenta satisfy $P_1^2=M_1^{*2}$, and $P_2^2=M_2^{*2}$, while $ k_{\nu_\ell}^2=0$ and $k_\ell^2=m_\ell^2$.

Collect the six real transition form factors into $\boldsymbol{\mathcal F}
    =
    \left(
        f_1,
        f_2,
        f_3,
        g_1,
        g_2,
        g_3
    \right)^T$. The form factors are taken to be real within the prescriptions used
in the present calculation.

The exact spin-summed squared amplitude can then be written as
\begin{equation}
    \sum_{\rm spins}
    \left|
        \mathcal M_{12\ell}
    \right|^2
    =
    32G_F^2C_{12}^2
    \,
    \boldsymbol{\mathcal F}^{T}
    \mathsf R
    \boldsymbol{\mathcal F},
    \label{eq:app_M2}
\end{equation}
where $\mathsf R$ is a real symmetric $6\times6$ matrix.

The vacuum masses $M_1$ and $M_2$ appearing below originate from the
normalization of the tensor and induced-scalar/pseudoscalar terms in
Equation~\eqref{eq:vertex}, whereas $M_1^*$ and $M_2^*$ denote the
corresponding in-medium effective masses.

Expanding Equation~\eqref{eq:app_M2},
\begin{align}
\boldsymbol{\mathcal F}^{T}
\mathsf R
\boldsymbol{\mathcal F}
={}&
R_{f_1f_1}f_1^2
+
R_{f_2f_2}f_2^2
+
R_{f_3f_3}f_3^2
+
R_{g_1g_1}g_1^2
+
R_{g_2g_2}g_2^2
+
R_{g_3g_3}g_3^2
\nonumber\\
&+
2R_{f_1f_2}f_1f_2
+
2R_{f_1f_3}f_1f_3
+
2R_{f_2f_3}f_2f_3
\nonumber\\
&+
2R_{g_1g_2}g_1g_2
+
2R_{g_1g_3}g_1g_3
+
2R_{g_2g_3}g_2g_3
\nonumber\\
&+
2\sum_{i,j=1}^{3}
R_{f_i g_j}f_i g_j .
\label{eq:app_R_expanded}
\end{align}

The independent vector-vector coefficients are
\begin{align}
R_{f_1f_1}
={}&
A_2B_1+A_1B_2-CM_1^*M_2^*,
\label{eq:Rf1f1}
\\
R_{f_1f_2}
={}&
\frac{1}{2(M_1+M_2)}
\Big[
2A_2B_2M_1^*
+2A_1B_1M_2^*
-A_2B_1(M_1^*+M_2^*)
-A_1B_2(M_1^*+M_2^*)
+C(M_1^*+M_2^*)
(M_1^*M_2^*-S)
\Big],
\label{eq:Kf1f2}
\\
R_{f_1f_3}
={}&
\frac{1}{2M_1}
\Big[
-2A_2B_2M_1^*
+A_2B_1(M_1^*-M_2^*)
+A_1B_2(M_1^*-M_2^*)
+2A_1B_1M_2^*
-C(M_1^*-M_2^*)
(M_1^*M_2^*+S)
\Big],
\label{eq:Rf1f3}
\\
R_{f_2f_2}
={}&
\frac{1}{2(M_1+M_2)^2}
\Big\{
-2A_1M_2^*
\left[
B_1M_1^*-B_2M_1^*+2B_1M_2^*
\right]
-C\left(M_1^{*2}+M_2^{*2}-2S\right)
\left(M_1^*M_2^*-S\right)
+2A_1(B_1+B_2)S
\nonumber\\
&\hspace{0.8cm}
+2A_2
\left[
B_1M_1^*M_2^*
-B_2M_1^*(2M_1^*+M_2^*)
+(B_1+B_2)S
\right]
\Big\},
\label{eq:Rf2f2}
\\
R_{f_2f_3}
={}&
\frac{1}{2M_1(M_1+M_2)}
\Big[
A_2B_1(-M_1^{*2}+M_2^{*2})
+A_1B_2(-M_1^{*2}+M_2^{*2})
+2A_2B_2(M_1^{*2}-S)
+2A_1B_1(-M_2^{*2}+S)
\Big],
\label{eq:Rf2f3}
\\
R_{f_3f_3}
={}&
-\frac{
\left[
-2(A_1-A_2)(B_1-B_2)
+C(M_1^{*2}+M_2^{*2}-2S)
\right]
(M_1^*M_2^*+S)
}{
2M_1^2
}.
\label{eq:Rf3f3}
\end{align}

The independent axial-axial coefficients are
\begin{align}
R_{g_1g_1}
={}&
A_2B_1+A_1B_2+CM_1^*M_2^*,
\label{eq:Rg1g1}
\\
R_{g_1g_2}
={}&
\frac{1}{2(M_1+M_2)}
\Big[
-2A_2 B_2M_1^*
+A_2 B_1(M_1^*-M_2^*) 
+A_1B_2(M_1^*-M_2^*)
+2A_1B_1M_2^*
+C(M_1^*-M_2^*)
(M_1^*M_2^*+S)
\Big],
\label{eq:Rg1g2}
\\
R_{g_1g_3}
={}&
-\frac{1}{2M_1}
\Big[
-2A_2B_2M_1^*
-2A_1B_1M_2^*
+A_2B_1(M_1^*+M_2^*)
+A_1B_2(M_1^*+M_2^*)
+C(M_1^*+M_2^*)
(M_1^*M_2^*-S)
\Big],
\label{eq:Rg1g3}
\\
R_{g_2g_2}
={}&
\frac{1}{2(M_1+M_2)^2}
\Big\{
-2A_2M_1^*
\left[
2B_2M_1^*+B_1M_2^*-B_2M_2^*
\right]
+2A_2(B_1+B_2)S
+C(M_1^{*2}+M_2^{*2}-2S)
(M_1^*M_2^*+S)
\nonumber\\
&\hspace{0.8cm}
+2A_1
\left[
M_2^*(B_1M_1^*-B_2M_1^*-2B_1M_2^*)
+(B_1+B_2)S
\right]
\Big\},
\label{eq:Rg2g2}
\\
R_{g_2g_3}
={}&
\frac{1}{2M_1(M_1+M_2)}
\Big[
A_2B_1(-M_1^{*2}+M_2^{*2})
+A_1B_2(-M_1^{*2}+M_2^{*2})
+2A_2B_2(M_1^{*2}-S)
+2A_1B_1(-M_2^{*2}+S)
\Big],
\label{eq:Rg2g3}
\\
R_{g_3g_3}
={}&
\frac{
\left[
-2(A_1-A_2)(B_1-B_2)
+C(M_1^{*2}+M_2^{*2}-2S)
\right]
(M_1^*M_2^*-S)
}{
2M_1^2
}.
\label{eq:Rg3g3}
\end{align}

The nonvanishing vector--axial interference coefficients are
\begin{align}
R_{f_1g_1}
={}&
-A_2B_1+A_1B_2,
\label{eq:Rf1g1}
\\
R_{f_1g_2}
={}&
-\frac{
(A_2B_1-A_1B_2)
(M_1^*-M_2^*)
}{
M_1+M_2
},
\label{eq:Rf1g2}
\\
R_{f_2g_1}
={}&
\frac{
(A_2B_1-A_1B_2)
(M_1^*+M_2^*)
}{
M_1+M_2
},
\label{eq:Rf2g1}
\\
R_{f_2g_2}
={}&
\frac{
(A_2B_1-A_1B_2)
(M_1^*-M_2^*)
(M_1^*+M_2^*)
}{
(M_1+M_2)^2
}.
\label{eq:Rf2g2}
\end{align}
The remaining vector--axial coefficients vanish,
\begin{equation}
R_{f_1g_3}
=
R_{f_2g_3}
=
R_{f_3g_1}
=
R_{f_3g_2}
=
R_{f_3g_3}
=
0.
\label{eq:Rfg_zero}
\end{equation}

\subsection{Strongly degenerate angular reduction}
\label{app:degenerate_reduction}

We now specialize the invariant decomposition to the strongly
degenerate regime used in the emissivity calculation. We choose the
$z$ axis along the momentum of the initial baryon,
$\boldsymbol{p}_1=p_1\hat{\boldsymbol z}$. Following
\citet{Leinson:2002bv},
\begin{align}
    P_1^\mu
    &=
    \left(
        \epsilon_1,
        0,
        0,
       |\boldsymbol{p}_1|
    \right),
    \label{eq:app_P1_frame}
    \\
    k_{\nu_\ell}^\mu
    &=
    \omega_{\nu_\ell}
    \left(
        1,
        \sin\theta_{\nu_\ell},
        0,
        \cos\theta_{\nu_\ell}
    \right),
    \label{eq:app_knu_frame}
    \\
    k_\ell^\mu
    &=
    \left(
        \omega_\ell,
        |\boldsymbol{k}_\ell|
        \sin\theta_\ell
        \cos\phi_\ell,
        |\boldsymbol{k}_\ell|
        \sin\theta_\ell
        \sin\phi_\ell,
        |\boldsymbol{k}_\ell|
        \cos\theta_\ell
    \right).
    \label{eq:app_kl_frame}
\end{align}

The baryons and charged lepton lie within $\mathcal{O}(T)$ of their respective
Fermi surfaces, whereas the emitted antineutrino carries energy and
momentum of order $T$. To leading order in $T/\mu_i$, the
antineutrino momentum may therefore be neglected when fixing the
smooth Fermi-surface three-momentum geometry, $\boldsymbol{p}_2
    \simeq
    \boldsymbol{p}_1
    -
    \boldsymbol{k}_\ell$.

The antineutrino four-momentum is nevertheless retained in the
matrix element and in the thermal phase-space integral. At this order,
\begin{equation}
    P_2^\mu
    \simeq
    \left(
        \epsilon_2,
        -|\boldsymbol{k}_\ell|
        \sin\theta_\ell
        \cos\phi_\ell,
        -|\boldsymbol{k}_\ell|
        \sin\theta_\ell
        \sin\phi_\ell,
        |\boldsymbol{p}_1|-|\boldsymbol{k}_\ell|\cos\theta_\ell
    \right).
    \label{eq:app_P2_frame}
\end{equation}
The magnitude of the final-baryon momentum satisfies $|\boldsymbol{p}_2|^2
    =
    |\boldsymbol{p}_1|^2+|\boldsymbol{k}_\ell|^2
    -
    2|\boldsymbol{p}_1||\boldsymbol{k}_\ell|\cos\theta_\ell$,

so that
\begin{equation}
    \cos\theta_\ell
    =
    \frac{
        |\boldsymbol{p}_1|^2+|\boldsymbol{k}_\ell|^2-|\boldsymbol{p}_2|^2
    }{
        2|\boldsymbol{p}_1||\boldsymbol{k}_\ell|
    }.
    \label{eq:app_costheta_l}
\end{equation}

Using Equation~\eqref{eq:app_costheta_l}, the invariants that do not
contain the antineutrino momentum reduce to
\begin{align}
    S
    &=
    \epsilon_1\epsilon_2
    +
    \frac{
        |\boldsymbol{k}_\ell|^2-|\boldsymbol{p}_1|^2-|\boldsymbol{p}_2|^2
    }{2},
    \label{eq:app_S_reduced}
    \\
    B_1
    &=
    \epsilon_1\omega_\ell
    -
    \frac{
        |\boldsymbol{p}_1|^2+|\boldsymbol{k}_\ell|^2-|\boldsymbol{p}_2|^2
    }{2},
    \label{eq:app_B1_reduced}
    \\
    B_2
    &=
    \epsilon_2\omega_\ell
    +
    \frac{
        |\boldsymbol{k}_\ell|^2-|\boldsymbol{p}_1|^2+|\boldsymbol{p}_2|^2
    }{2}.
    \label{eq:app_B2_reduced}
\end{align}

Define the angle $\gamma$ between the charged-lepton and
antineutrino momenta through
\begin{equation}
    \cos\gamma
    =
    \hat{\boldsymbol{k}}_{\nu_\ell}
    \cdot
    \hat{\boldsymbol{k}}_\ell
    =
    \sin\theta_{\nu_\ell}
    \sin\theta_\ell
    \cos\phi_\ell
    +
    \cos\theta_{\nu_\ell}
    \cos\theta_\ell.
    \label{eq:app_cosgamma}
\end{equation}
The remaining three invariants are
\begin{align}
    A_1
    &=
    \omega_{\nu_\ell}
    \left(
        \epsilon_1
        -
        |\boldsymbol{p}_1|\cos\theta_{\nu_\ell}
    \right),
    \label{eq:app_A1_angular}
    \\
    A_2
    &\simeq
    \omega_{\nu_\ell}
    \left(
        \epsilon_2
        -
        |\boldsymbol{p}_1|\cos\theta_{\nu_\ell}
        +
        |\boldsymbol{k}_\ell|\cos\gamma
    \right),
    \label{eq:app_A2_angular}
    \\
    C
    &=
    \omega_{\nu_\ell}
    \left(
        \omega_\ell
        -
        |\boldsymbol{k}_\ell|\cos\gamma
    \right).
    \label{eq:app_C_angular}
\end{align}
Thus, $A_1$, $A_2$, and $C$ each contain one overall factor of the
antineutrino energy $\omega_{\nu_\ell}$. Since every nonvanishing
coefficient $R_{XY}$ contains exactly one factor from the set
$\{A_1,A_2,C\}$, the complete quadratic form is linear in
$\omega_{\nu_\ell}$.

For fixed charged-lepton direction, define the normalized
antineutrino angular average
\begin{equation}
    \left\langle X\right\rangle_{\nu_\ell}
    \equiv
    \frac{1}{4\pi}
    \int d\Omega_{\nu_\ell}\,X.
    \label{eq:app_nu_average}
\end{equation}
Since $\left\langle
        \cos\theta_{\nu_\ell}
    \right\rangle_{\nu_\ell}
    =
    \left\langle
        \cos\gamma
    \right\rangle_{\nu_\ell}
    =
    0$, we obtain
\begin{equation}
    \left\langle A_1\right\rangle_{\nu_\ell}
    =
    \omega_{\nu_\ell}\epsilon_1,
    \qquad
    \left\langle A_2\right\rangle_{\nu_\ell}
    =
    \omega_{\nu_\ell}\epsilon_2,
    \qquad
    \left\langle C\right\rangle_{\nu_\ell}
    =
    \omega_{\nu_\ell}\omega_\ell.
    \label{eq:app_ABC_average}
\end{equation}

We then evaluate the smooth quantities at their Fermi surfaces,
\begin{equation}
    |\boldsymbol{p}_1|=k_{F1},
    \qquad
    |\boldsymbol{p}_2|=k_{F2},
    \qquad
    |\boldsymbol{k}_\ell|=k_{F\ell},
    \qquad
    \epsilon_i=\epsilon_{Fi},
    \qquad
    \omega_\ell=\mu_\ell.
    \label{eq:app_FS_substitutions}
\end{equation}
The reduced invariants used in the numerical calculation are
therefore
\begin{align}
    S
    &=
    \epsilon_{F1}\epsilon_{F2}
    +
    \frac{
        k_{F\ell}^2-k_{F1}^2-k_{F2}^2
    }{2},
    \label{eq:app_S_final}
    \\
    B_1
    &=
    \epsilon_{F1}\mu_\ell
    -
    \frac{
        k_{F1}^2+k_{F\ell}^2-k_{F2}^2
    }{2},
    \label{eq:app_B1_final}
    \\
    B_2
    &=
    \epsilon_{F2}\mu_\ell
    +
    \frac{
        k_{F\ell}^2-k_{F1}^2+k_{F2}^2
    }{2},
    \label{eq:app_B2_final}
\end{align}
together with
\begin{equation}
    A_1
    =
    \omega_{\nu_\ell}\epsilon_{F1},
    \qquad
    A_2
    =
    \omega_{\nu_\ell}\epsilon_{F2},
    \qquad
    C
    =
    \omega_{\nu_\ell}\mu_\ell.
    \label{eq:app_ABC_final}
\end{equation}

For the momentum-dependent form-factor prescription adopted in the
main text, the corresponding kinetic Fermi-surface transfer is $q_F^2
    =
    \left(
        \epsilon_{F1}-\epsilon_{F2}
    \right)^2
    -
    k_{F\ell}^2$, equivalent to Equation~\eqref{eq:q2_fermi_surface}.

After the angular average,
Equation~\eqref{eq:app_M2} takes the form
\begin{equation}
    \sum_{\rm spins}
    \left|
        \mathcal M_{12\ell}
    \right|^2
    \simeq
    32G_F^2C_{12}^2
    \omega_{\nu_\ell}
    \mathcal A_{12\ell},
    \label{eq:app_M2_omega}
\end{equation}
where $\mathcal A_{12\ell}$ is independent of $\omega_{\nu_\ell}$ at the
order retained.

\subsection{Thermal integral and final emissivity}
\label{app:thermal_reduction}

The antineutrino phase-space measure is
\begin{equation}
    d\Pi_{\nu_\ell}
    =
    \frac{
        \omega_{\nu_\ell}
        \,d\omega_{\nu_\ell}
        \,d\Omega_{\nu_\ell}
    }{
        2(2\pi)^3
    }.
    \label{eq:app_nu_measure}
\end{equation}

The emissivity contains one additional factor of
$\omega_{\nu_\ell}$ for the energy carried by the emitted
antineutrino. Together with Equation~\eqref{eq:app_M2_omega}, for which the
spin-summed squared amplitude is linear in
$\omega_{\nu_\ell}$, the antineutrino-energy dependence of the
integrand is therefore
$\omega_{\nu_\ell}^{3}d\omega_{\nu_\ell}$.

To make the remaining thermal reduction explicit, we introduce
the excitation energies measured relative to the corresponding
chemical potentials,
\begin{equation}
\xi_1
=
\epsilon_1+U_1-\mu_1,
\qquad
\xi_2
=
\epsilon_2+U_2-\mu_2,
\qquad
\xi_\ell
=
\omega_\ell-\mu_\ell .
\end{equation}
Using the Fermi--Dirac distributions defined in Equation~\eqref{eq:Fermi-Dirac},
we may therefore write $f_1(\epsilon_1)=n_F(\xi_1)$, $f_2(\epsilon_2)=n_F(\xi_2)$, and $f_\ell(\omega_\ell)=n_F(\xi_\ell)$, where $n_F(x)=\frac{1}{e^{x/T}+1}$.

The temporal component of the four-dimensional
energy--momentum conserving delta function in Equation~\eqref{eq:app_full_Q}
imposes $\epsilon_1+U_1-\epsilon_2-U_2
-\omega_\ell-\omega_{\nu_\ell}=0$.

Using beta equilibrium, $\mu_1-\mu_2=\mu_\ell$, this relation becomes 
\begin{equation}
    \xi_1-\xi_2-\xi_\ell-\omega_{\nu_\ell}=0.
\end{equation}

It is convenient to define $\xi\equiv\xi_1$
and $\omega'\equiv\xi_1-\xi_2$.
The second definition immediately gives $\xi_2=\xi-\omega'$. On the other hand, the temporal component of the
energy--momentum conserving delta function implies $\epsilon_1+U_1
=
\epsilon_2+U_2+\omega_\ell+\omega_{\nu_\ell}$. Using $\epsilon_i+U_i=\mu_i+\xi_i$, and $\omega_\ell=\mu_\ell+\xi_\ell$, together with beta equilibrium, this becomes $\xi_1-\xi_2
=
\xi_\ell+\omega_{\nu_\ell}$, and hence
\begin{equation}
\xi_\ell=\omega'-\omega_{\nu_\ell}.
\end{equation}
The occupation and blocking factors therefore reduce to
\begin{equation}
f_1(\epsilon_1)
\left[1-f_2(\epsilon_2)\right]
\left[1-f_\ell(\omega_\ell)\right]
=
n_F(\xi)
\left[1-n_F(\xi-\omega')\right]
\left[1-n_F(\omega'-\omega_{\nu_\ell})\right].
\end{equation}

The integration over the initial-baryon excitation energy can
then be performed analytically using
\begin{equation}
\int_{-\infty}^{\infty} d\xi\,
n_F(\xi)
\left[1-n_F(\xi-\omega')\right]
=
\frac{\omega'}{\exp(\omega'/T)-1}.
\end{equation}
Furthermore,
\begin{equation}
1-n_F(\omega'-\omega_{\nu_\ell})
=
\frac{1}
{1+\exp[(\omega_{\nu_\ell}-\omega')/T]}.
\end{equation}
Within the strongly degenerate approximation introduced above,
the remaining thermal integral is
\begin{align}
&\int_0^\infty d\omega_{\nu_\ell}\,
\omega_{\nu_\ell}^{3}
\int_{-\infty}^{\infty}d\omega'
\int_{-\infty}^{\infty}d\xi\,
n_F(\xi)
\left[1-n_F(\xi-\omega')\right]
\left[1-n_F(\omega'-\omega_{\nu_\ell})\right]
\nonumber\\
&=
\int_{-\infty}^{\infty}d\omega'\,
\frac{\omega'}{\exp(\omega'/T)-1}
\int_0^\infty d\omega_{\nu_\ell}\,
\frac{\omega_{\nu_\ell}^{3}}
{1+\exp[(\omega_{\nu_\ell}-\omega')/T]} =
\frac{457\pi^6}{5040}\,T^6.
\end{align}

The spatial component of the same four-dimensional delta function
enforces
\begin{equation}
\mathbf{p}_1-\mathbf{p}_2-\mathbf{k}_\ell
-\mathbf{k}_{\nu_\ell}=0 .
\end{equation}
To leading order in the strongly degenerate limit,
$|\mathbf{k}_{\nu_\ell}|\sim T$ may be neglected when fixing the
Fermi-surface geometry, giving
$\mathbf{p}_2\simeq\mathbf{p}_1-\mathbf{k}_\ell$.
At the Fermi surfaces this implies
\begin{equation}
k_{F2}^2=k_{F1}^2+k_{F\ell}^2
-2k_{F1}k_{F\ell}\cos\theta_\ell ,
\end{equation}
and hence

\begin{equation}
    \cos\theta_\ell^{\,0}
    =
    \frac{
        k_{F1}^2+k_{F\ell}^2-k_{F2}^2
    }{
        2k_{F1}k_{F\ell}
    }.
    \label{eq:app_cos_theta_l_final}
\end{equation}
The condition $-1
    \leq
    \cos\theta_\ell^{\,0}
    \leq
    1$ is equivalent to the requirement that the three Fermi momenta form a
triangle and therefore produces the threshold factor
$\Theta_{12\ell}$ of Equation~\eqref{eq:triangle_factor}.

Combining the angular reduction with the thermal integral gives
\begin{equation}
    Q_{12\ell}
    \simeq
    \frac{457\pi}{10080}
    G_F^2C_{12}^2
    T^6
    \Theta_{12\ell}
    \mathcal A_{12\ell}
    \label{eq:app_final_emissivity}
\end{equation}
which is the result quoted in the main text.

\section{Equation-of-state dependence}
\label{app:eos_dependence}
\begin{figure}
\centering

\begin{minipage}[t]{0.48\textwidth}
    \vspace{0pt}
    \centering
    \includegraphics[width=\linewidth]
    {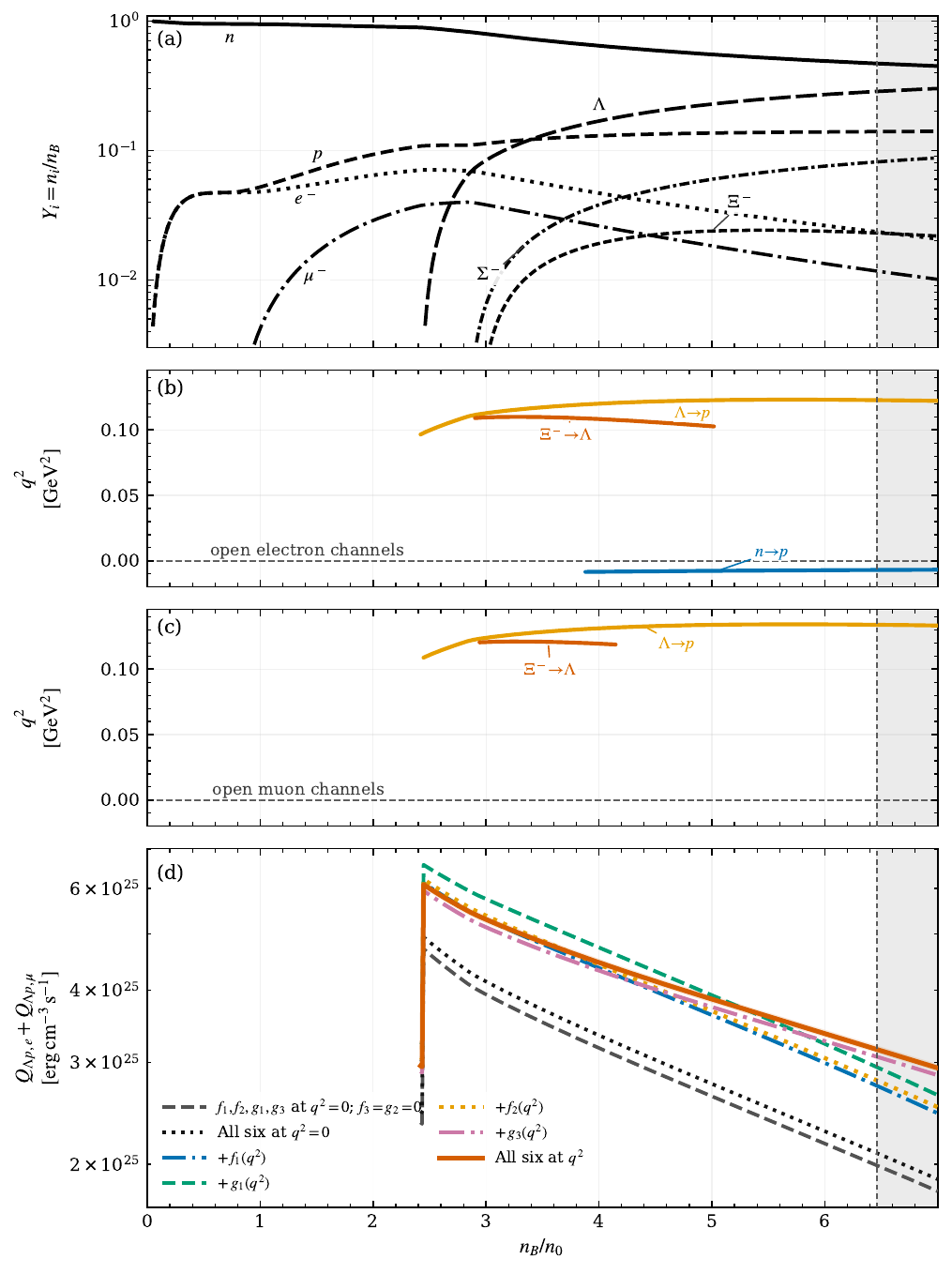}
\end{minipage}
\hfill
\begin{minipage}[t]{0.48\textwidth}
    \vspace{0pt}
    \centering

    \includegraphics[width=\linewidth]
    {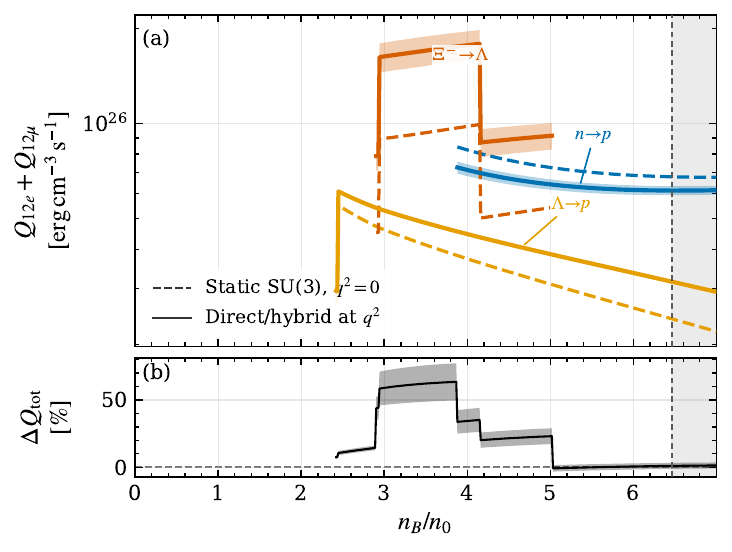}

    \vspace{0.5em}

    \includegraphics[width=\linewidth]
    {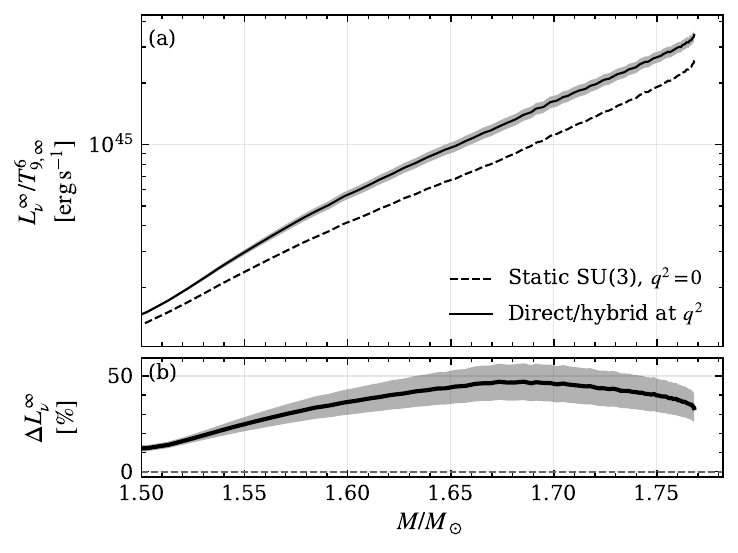}
\end{minipage}

\caption{
FSU2R sensitivity test corresponding to
Figures~\ref{fig:density_kinematics}--\ref{fig:luminosity}.
Left: composition, kinetic momentum transfer, and progressive
$\Lambda\to p$ form-factor dependence.
Upper right: channel-resolved direct-Urca emissivities and
$\Delta Q_{\rm tot}$.
Lower right: redshifted neutrino luminosity and
$\Delta L_\nu^\infty$ along the stable stellar sequence.
The qualitative enhancement of the hyperonic weak-vertex
sensitivity persists, while its quantitative magnitude is EOS
dependent.
}
\label{fig:fsu2r_summary}
\end{figure}

To assess the dependence of the results on the dense-matter model,
we repeated the analysis using the hyperonic FSU2R parametrization
of \citet{Tolos:2016hhl}. FSU2R and FSU2H belong to the same
nonlinear RMF framework and use the same hyperon-potential
calibration,
$U_\Lambda^{(N)}=-28$ MeV,
$U_\Sigma^{(N)}=+30$ MeV, and
$U_\Xi^{(N)}=-18$ MeV,
but differ appreciably in their high-density behavior. The $\Lambda$ potential is comparatively well constrained by
hypernuclear data, whereas the $\Sigma$ and especially $\Xi$
potentials remain less precisely determined. Variations of these
interactions constitute an additional systematic uncertainty not
explored here.
This comparison is therefore intended as a sensitivity check rather
than as a systematic study of EOS or hyperon-interaction
uncertainties.

Figure~\ref{fig:fsu2r_summary} summarizes the corresponding
FSU2R results. The qualitative kinematic picture remains unchanged:
$n\to p$ probes $q_F^2\simeq0$, whereas the hyperonic
$\Lambda\to p$ and $\Xi^-\to\Lambda$ channels probe timelike
transfers of order $0.1~{\rm GeV}^2$ and are therefore substantially
more sensitive to the momentum dependence of the weak vertex.
The quantitative effect, however, changes with the EOS. Both the
local emissivity correction and the integrated luminosity correction
differ noticeably from the FSU2H results. This EOS dependence is
expected because the composition, mean fields, direct-Urca
thresholds, and stellar density profiles are all EOS dependent.
A systematic investigation of this dependence is beyond the scope
of the present work.

\newpage

\bibliography{biblio}{}
\bibliographystyle{aasjournal}

\listofchanges

\end{document}